\documentclass[twocolumn]{raa}
 
\usepackage{graphicx,times}             
\usepackage{natbib}
\usepackage{amssymb,amsmath}
\usepackage{mathrsfs}            
\usepackage{amsmath}
\usepackage{graphicx,graphics}	% Including figure files
\usepackage[T1]{fontenc}
\usepackage{pbox}
\usepackage{booktabs}
\usepackage{tabularx}
\usepackage[utf8]{inputenc}
\usepackage{subcaption}
\usepackage{enumitem}
\usepackage{soul}
\usepackage{xcolor}

\bibpunct{(}{)}{;}{a}{}{,}
 
\usepackage[pagebackref=true]{hyperref}
\soulregister\autoref7
\soulregister\ref7
\soulregister\cite7

\def\apjs{ApJS}

\def\aap{A\&A}

\def\apjs{ApJS}
\def\apjl{ApJ Letters}

\begin{document}

\titlerunning{Effect of AGN proximity on star formation}

\authorrunning{A. Das \& B. Pandey}

\title{Triggering and quenching in the shadow of AGN: How does AGN proximity affect star formation in the EAGLE simulation?}

 \volnopage{Vol.0 (20xx) No.0, 000--000}     
   \setcounter{page}{1}           

 \author{Apashanka Das \inst{1} \and Biswajit Pandey \inst{2} }
 \institute{Harish-Chandra Research Institute, HBNI, Chhatnag Road, Jhunsi, Allahabad, 211109, India
             {\it a.das.cosmo@gmail.com}\\
        \and
	     {Department of Physics, Visva-Bharati University, Santiniketan, 
	     Birbhum, 731235, India {\it biswap@visva-bharati.ac.in}}\\
   }
       
\vs\no
   {\small Received 20xx month day; accepted 20xx month day}

   \abstract{Active galactic nuclei (AGN) inject vast amounts of energy into their surroundings and are widely recognised as key drivers of galaxy evolution through feedback processes. Although the effects of AGN feedback within host galaxies are well studied, the extent to which AGN influence star formation in neighbouring galaxies remains an open question. In this work, we use the EAGLE cosmological hydrodynamical simulation to investigate how AGN proximity modulates star formation in nearby star-forming galaxies (SFGs) on scales up to 2~Mpc. We employ a carefully constructed control sample matched in stellar mass and local density to isolate the environmental impact of AGN, and quantify deviations in star formation using an SFR offset relative to matched controls. Our analysis reveals a clear, non-uniform response to AGN proximity: low-mass galaxies exhibit systematically different star formation responses than their high-mass counterparts, with gas-rich systems frequently showing mild star formation enhancement, consistent with triggering via AGN-driven compression or turbulence, while gas-poor systems preferentially exhibit suppressed star formation, indicative of thermal heating inhibiting gas cooling. Cold gas availability is found to be more strongly associated with this modulation than stellar mass alone. These signatures are strongest within $\sim 1$~Mpc of AGN hosts but persist out to 2~Mpc, demonstrating that AGN feedback operates well beyond the host halo. Linear fits reveal that these signatures are strongest within $\sim1$~Mpc of AGN hosts and gradually weaken with increasing separation, while remaining detectable out to 2~Mpc, demonstrating that AGN feedback operates well beyond the host halo. Together, our results establish AGN as non-local regulators of galaxy evolution, capable of shaping star formation in neighbouring galaxies through extended feedback processes that link black hole accretion physics to the large-scale cosmic environment. }
   
\date{\today} 
\maketitle

\section{Introduction}

Active galactic nuclei (AGN) represent extreme phases in the life cycle of galaxies, during which supermassive black holes (SMBHs) at galactic centres accrete matter at high rates and release enormous amounts of energy. These episodes can generate bolometric luminosities approaching $10^{47}$-$10^{48}$~erg~s$^{-1}$ across the electromagnetic spectrum, from hard X-rays to radio wavelengths \citep{fabian99, woo02}. The energy liberated during AGN activity is transferred to the surrounding gas through a combination of radiation pressure, relativistic jets, and fast outflows, giving rise to what is broadly termed AGN feedback. This feedback can profoundly alter the thermodynamic state of the interstellar and circumgalactic medium by heating, displacing, or expelling gas, thereby regulating both star formation and black hole growth \citep{kawata05, wagner13, morganti17, baron18, santoro20}. Consequently, AGN feedback has become an essential ingredient of contemporary galaxy formation models, particularly in explaining the suppression of star formation in massive systems and the emergence of the observed bimodality between star-forming and quiescent galaxies \citep{springel05, somerville08, kormendy13, heckman14, harrison17, baldry04, das25}.

Although SMBHs are now believed to be nearly ubiquitous in massive galaxies, AGN activity itself is intermittent, implying that accretion episodes are regulated by a combination of internal and external factors. Numerous studies have demonstrated that properties such as gas availability, bulge structure, and central potential depth play a critical role in enabling or limiting black hole accretion \citep{ruffa19, shangguan20, ellison21, sampaio23}. At the same time, growing observational evidence suggests that environment also matters. Elevated AGN fractions have been reported in galaxies residing within groups and cluster outskirts \citep{lopes17, aird21}, while other works find enhanced AGN activity in relatively underdense regions, including cosmic voids \citep{ceccarelli21}. More recently, \citet{banerjee25} showed that, even at fixed stellar mass, AGN host galaxies differ systematically from purely star-forming systems in colour, star formation rate, morphology, and stellar population age across a wide range of environments. Taken together, these results point to a complex interplay between internal galaxy structure, assembly history, and environment in governing when and where AGN are triggered.

Beyond their influence on host galaxies, AGN may also affect their surroundings on much larger scales. Clustering analyses reveal that AGN are more strongly clustered than typical star-forming galaxies \citep{gilli09, mandelbaum09, donoso14, hale18, singh23}, and the likelihood of AGN activity increases in the presence of close galactic neighbours \citep{satyapal14, zhang21}. These trends suggest that AGN-related processes may operate beyond individual haloes, potentially influencing the intergalactic medium and nearby galaxies. Striking evidence for such extended influence is provided by radio-loud AGN, including quasars and giant radio galaxies (GRGs), whose jets can propagate over megaparsec scales \citep{hardcastle19, oei23}. Observations indicate that GRGs commonly span $0.7$ -- $3.5$~Mpc \citep{dhabade20}, while extreme systems such as Alcyoneus exhibit radio structures extending beyond $5$~Mpc \citep{oei22}. These enormous outflows inject energy and momentum into the surrounding medium, redistribute magnetic fields, and may either suppress or stimulate star formation in neighbouring galaxies. Such observations motivate a critical question: can AGN activity leave a measurable, non-local imprint on the star formation properties of galaxies well beyond the host halo?

Addressing this question requires a framework that combines physical realism with sufficient statistical power to probe environmental effects across a broad range of scales. The EAGLE (Evolution and Assembly of GaLaxies and their Environments) cosmological hydrodynamical simulation provides such a framework. Its large cosmological volumes, high-resolution hydrodynamics, and carefully calibrated subgrid prescriptions for star formation and AGN feedback enable it to reproduce a wide array of observed galaxy properties \citep{schaye15, furlong15, trayford15, mcalpine16, katsianis17, vellis17, baes20}. The resulting galaxy catalogues offer a statistically robust platform for isolating subtle environmental trends in galaxy evolution.

In this work, we use the EAGLE simulation to investigate whether proximity to AGN systematically modulates the star formation activity of neighbouring galaxies. We focus on star-forming galaxies located within 2~Mpc of AGN hosts and compare their properties to those of a carefully constructed control sample matched in stellar mass and local density. By quantifying deviations in star formation rate relative to these controls, we aim to isolate the influence of AGN proximity from intrinsic galaxy properties and large-scale environmental effects. This approach allows us to explore the non-local footprint of AGN feedback and assess its role in shaping galaxy evolution beyond the confines of individual host haloes.

\section{Data and Method of Analysis}           

\subsection{EAGLE simulation data}
\label{sec:data}

Our analysis is based on galaxy catalogues drawn from the publicly released EAGLE (Evolution and Assembly of GaLaxies and their Environments) cosmological hydrodynamical simulation suite \citep{schaye15, crain15, mcalpine16}. The EAGLE simulations follow the coupled evolution of dark matter and baryons within periodic, cubic comoving volumes spanning side lengths from 25 to 100~Mpc, beginning at an initial redshift of $z=127$ and evolving to the present epoch. The simulations adopt a flat $\Lambda$CDM cosmology consistent with \textit{Planck} results \citep{planck14}, characterized by $\Omega_{\Lambda}=0.693$, $\Omega_{\mathrm{m}}=0.307$, $\Omega_{\mathrm{b}}=0.04825$, and a Hubble constant of $H_0 = 67.77~\mathrm{km\,s^{-1}\,Mpc^{-1}}$.

In this work, we focus on the flagship reference model \texttt{Ref-L0100N1504}, which simulates galaxy formation within a comoving volume of $(100~\mathrm{Mpc})^3$ using a total of $2 \times 1504^3$ particles. The initial mass resolution of this model corresponds to $1.81 \times 10^{6}\,M_{\text{sun}}$ for baryonic particles and $9.70 \times 10^{6}\,M_{\text{sun}}$ for dark matter particles. We restrict our analysis to the snapshot at redshift $z=0$, enabling a detailed examination of galaxy properties in the local Universe. To ensure that only physically meaningful and numerically robust systems are included, we exclude artefacts identified by the \texttt{Spurious} flag, thereby removing objects with unreliable stellar, metallicity, or black hole properties.

Galaxy physical properties, including stellar mass and star formation rate (SFR), are obtained from the \texttt{Ref-L0100N1504-Aperture} and \texttt{Ref-L0100N1504-Magnitudes} catalogues \citep{trayford15, doi10}. These quantities are measured within a three-dimensional spherical aperture of radius 30~physical~kpc, centred on the location of each galaxy’s minimum gravitational potential. This aperture choice has been shown to yield galaxy properties that are directly comparable to observational measurements \citep{mcalpine16}. To ensure adequate numerical resolution and reliable estimates of star formation activity, we further impose a lower stellar mass cut of $\log(M_{\star}/M_{\text{sun}}) \geq 8.3$, corresponding to a minimum of $110$ baryonic particles per galaxy.

\subsection{Local density estimation}

To quantify the immediate environment of each galaxy, we employ a nearest-neighbour based local density estimator following the formalism introduced by \citet{casertano85}. Specifically, the local galaxy density, $\eta_k$, is defined as
\begin{equation}
\eta_k = \frac{k-1}{V(r_k)},
\end{equation}
where $r_k$ denotes the three-dimensional distance to the $k^{\mathrm{th}}$ nearest neighbouring galaxy and $V(r_k)$ is the volume of a sphere of radius $r_k$. This estimator provides a robust measure of the surrounding galaxy number density on adaptive spatial scales.

In this study, we adopt $k=5$ and compute the fifth-nearest neighbour density, $\eta_5$, for each galaxy. This choice balances sensitivity to small-scale environments while mitigating shot noise associated with very local estimates. As the EAGLE simulations are defined within finite periodic volumes, special care is required for galaxies located near the simulation boundaries. To prevent systematic underestimation of local densities in such cases, we exclude galaxies whose distance to the nearest simulation boundary, $r_b$, is smaller than their fifth-nearest neighbour distance ($r_b < r_5$). This selection ensures that all density estimates are computed within fully sampled spherical volumes and are therefore reliable.

\subsection{Identification of star-forming and AGN host galaxies in \textsc{EAGLE}}
\label{subsec:sfg-agn-selection}

Galaxies hosting active galactic nuclei (AGN) are characterised by an additional, non-stellar source of radiation arising from accretion onto a central supermassive black hole. This accretion-powered emission spans the full electromagnetic spectrum and can dominate the total luminosity of the galaxy, in contrast to star-forming galaxies (SFGs), whose radiative output is primarily driven by stellar populations. AGN activity is also widely recognised as a key agent in regulating star formation by heating, expelling, or redistributing gas within and around galaxies, potentially driving the transition from active star formation to quiescence. In observations, these distinct energy sources are often separated using emission-line diagnostics \citep{baldwin81}. However, in simulations such as \textsc{EAGLE}, physically motivated quantities can be used to classify galaxies in a consistent and reproducible manner.

In this work, we identify AGN host galaxies based on their bolometric AGN luminosity, following the methodology adopted in previous \textsc{EAGLE} studies \citep{mcalpine20}. A galaxy is classified as AGN-active if its bolometric luminosity satisfies
\begin{equation}
L_{\mathrm{bol}} \geq 10^{43}\,\mathrm{erg\,s^{-1}}.
\end{equation}
The bolometric luminosity is computed as
\begin{equation}
L_{\mathrm{bol}} = \epsilon_r\,\dot{m}_{\mathrm{BH}}\,c^2,
\end{equation}
where $\dot{m}_{\mathrm{BH}}$ is the black hole accretion rate, $c$ is the speed of light, and $\epsilon_r = 0.1$ is the radiative efficiency of the accretion flow \citep{shakura73}. This prescription quantifies the fraction of rest-mass energy converted into radiation during accretion.

Star-forming galaxies are identified using a specific star formation rate threshold of
\begin{equation}
\mathrm{sSFR} > 0.01~\mathrm{Gyr}^{-1},
\end{equation}
consistent with the definition adopted in \citet{schaye15}. This criterion effectively separates actively star-forming systems from quiescent galaxies at $z=0$.

Applying these criteria to the $z=0$ galaxy catalogue yields a total of 29,734 galaxies, which we divide into four mutually exclusive classes based on their star formation and AGN activity. The classification scheme and corresponding sample sizes are summarised in Table~\ref{tab:galaxy_classification}.

\begin{table}[h!]
\centering
\caption{Galaxy classification at $z=0$ in the \textsc{EAGLE} simulation. The table lists the explicit selection criteria used to identify star-forming galaxies and AGN hosts, along with the resulting sample sizes.}
\label{tab:galaxy_classification}
\begin{tabular}{lcl}
\hline
\hline
Galaxy class & Selection criteria & Number \\
\hline
Star-forming (non-AGN) 
& $\mathrm{sSFR} > 0.01~\mathrm{Gyr}^{-1}$ and $L_{\mathrm{bol}} < 10^{43}\,\mathrm{erg\,s^{-1}}$ 
& 17,589 \\[4pt]

AGN (non-star-forming) 
& $\mathrm{sSFR} \leq 0.01~\mathrm{Gyr}^{-1}$ and $L_{\mathrm{bol}} \geq 10^{43}\,\mathrm{erg\,s^{-1}}$ 
& 85 \\[4pt]

Star-forming AGN 
& $\mathrm{sSFR} > 0.01~\mathrm{Gyr}^{-1}$ and $L_{\mathrm{bol}} \geq 10^{43}\,\mathrm{erg\,s^{-1}}$ 
& 189 \\[4pt]

Quiescent (non-AGN) 
& $\mathrm{sSFR} \leq 0.01~\mathrm{Gyr}^{-1}$ and $L_{\mathrm{bol}} < 10^{43}\,\mathrm{erg\,s^{-1}}$ 
& 11,871 \\
\hline
Total & -- & 29,734 \\
\hline
\hline
\end{tabular}
\end{table}

While AGN feedback is known to exert a strong influence on the evolution of host galaxies particularly through jet-driven heating of the circumgalactic medium in massive haloes, our focus in this study lies beyond the host itself. Specifically, we investigate whether the presence of an AGN can affect the star formation activity of neighbouring galaxies on megaparsec scales.

To this end, we construct two comparison samples of star-forming galaxies:
\begin{enumerate}[label=(\roman*)]
    \item star-forming galaxies located within a three-dimensional distance of 2~Mpc from an AGN host, and
    \item star-forming galaxies with no AGN present within 2~Mpc.
\end{enumerate}
Our AGN sample includes both purely AGN-hosting galaxies and galaxies that simultaneously exhibit AGN activity and ongoing star formation. This choice maximises the statistical power of the analysis while maintaining a physically well-defined AGN population. To isolate the effect of AGN proximity from other confounding factors, the two star-forming samples are subsequently matched in stellar mass and local density, as described in the following subsection.

\subsection{Constructing the control samples}
\label{control}

Both environment and stellar mass are known to exert a strong influence on the star formation properties of galaxies. Observationally, quiescent, red-sequence galaxies are preferentially found in dense environments, whereas actively star-forming systems dominate lower-density regions. A variety of environmental mechanisms such as ram-pressure stripping \citep{gunn72}, tidal harassment \citep{moore96}, strangulation \citep{balogh00}, and starvation \citep{larson80} operate most efficiently in groups and clusters, where they can remove or heat cold gas and thereby suppress star formation. Any attempt to isolate the effect of AGN proximity must therefore carefully control for environmental density.

Stellar mass provides an additional, independent axis of galaxy evolution. Numerous studies have shown that galaxies exhibit a marked transition in their physical properties around a characteristic stellar mass of $\sim3\times10^{10}\,M_{\text{sun}}$ \citep{kauffmann03}. Theoretically and in hydrodynamical simulations, this scale is associated with a dark matter halo mass of order $10^{12}\,M_{\text{sun}}$, above which stable virial shocks develop and gas accretion shifts from a cold to a hot mode \citep{binney04, birnboim03, dekel06, keres05, gabor10, gabor15}. Controlling for stellar mass is therefore essential to disentangle AGN-driven effects from mass-dependent trends in star formation.

Our AGN catalogue comprises a total of 274 galaxies, including 85 systems classified as purely AGN-dominated and 189 galaxies that host both AGN activity and ongoing star formation. These galaxies are used exclusively as AGN hosts and serve as the centres around which we examine potential environmental influence. Within a three-dimensional radius of 2~Mpc around these AGN hosts, we identify 2,157 star-forming galaxies that do not themselves host any AGN activity. This set of galaxies constitutes our AGN-vicinity (reference) sample. Star-forming AGN are explicitly excluded from this reference population to ensure that any observed modulation of star formation can be attributed to neighbouring AGN rather than internal nuclear activity.

To establish a robust baseline for comparison, we construct a control sample drawn from star-forming galaxies that lie beyond 2~Mpc of any AGN. For each reference galaxy, we select five control galaxies matched in stellar mass and local density, requiring agreement within $\pm0.1$~dex in stellar mass and $\pm0.095$~dex in local density. The environmental density tolerance of $\pm0.095$ dex was determined empirically to provide a sufficient number of suitable control galaxies while maintaining a close match in local environmental density. Although very similar to a tolerance of $\pm0.1$ dex, this value was found to provide the best balance between matching accuracy and control sample completeness. The statistical equivalence of the reference and control samples is verified using a Kolmogorov-Smirnov (K-S) test. In both stellar mass and local density, the null hypothesis that the two samples are drawn from the same parent distribution cannot be rejected at confidence levels exceeding 50\%, indicating that the matched samples are statistically indistinguishable in these properties. This matching procedure ensures that any systematic differences in star formation activity arise from AGN proximity rather than underlying mass or environmental biases.

While comparisons between the overall reference and control populations provide valuable insight, they can mask object-by-object variations. To capture these individual-level deviations, we quantify star formation offsets for each galaxy relative to its matched controls, following the methodology introduced by \citet{patton11}. These offset metrics form the basis of our subsequent analysis and enable a direct assessment of how star formation in individual galaxies responds to the presence of nearby AGN.

\subsection{$\Delta \mathrm{SFR}$ offsets}

To quantify how the star formation activity of galaxies responds to the presence of nearby AGN, we employ an offset-based framework inspired by \citet{patton11}. The central idea of this approach is to evaluate each galaxy relative to a small, well-matched comparison set, thereby minimizing the influence of global correlations with stellar mass and environment and isolating deviations potentially associated with AGN proximity.

For every star-forming galaxy $G$ in the AGN-vicinity (hereafter the reference galaxy), defined as having at least one AGN within a comoving distance of 2~Mpc, we construct an associated control set $\mathcal{C}_G = \{C_1, C_2, \ldots, C_5\}$. The five control galaxies are required to (i) be star forming, (ii) host no AGN within 2~Mpc, (iii) match the reference galaxy in stellar mass within $\pm0.1$~dex, and (iv) match in local density within $\pm0.095$~dex. This matching ensures that the control galaxies represent the expected star formation behaviour in the absence of nearby AGN, given the same mass and environmental conditions.

The star formation rate (SFR) offset of the reference galaxy is defined as the difference between its SFR and the mean SFR of its matched controls:
\begin{equation}
(\mathrm{SFR}\,\mathrm{offset})_G = \mathrm{SFR}_G - \left( \frac{1}{5} \sum_{i=1}^{5} \mathrm{SFR}_{C_i} \right),
\end{equation}
where $\mathrm{SFR}_{C_i}$ denotes the SFR of the $i$th control galaxy.

To provide a symmetric reference scale for this comparison, we also compute an internal offset for each control galaxy $C_j \in \mathcal{C}_G$ relative to the remaining four controls:
\begin{equation}
(\mathrm{SFR}\,\mathrm{offset})_{C_j} = \mathrm{SFR}_{C_j} - \left( \frac{1}{4} \sum_{\substack{i=1 \\ i \neq j}}^{5} \mathrm{SFR}_{C_i} \right).
\end{equation}
The mean of these five quantities defines the effective SFR offset of the control set:
\begin{equation}
\left\langle (\mathrm{SFR}\,\mathrm{offset})_{C_j} \right\rangle = \frac{1}{5} \sum_{j=1}^{5} (\mathrm{SFR}\,\mathrm{offset})_{C_j}.
\end{equation}

Finally, we define the differential star formation metric for the reference galaxy as
\begin{equation}
\Delta \mathrm{SFR}\,\mathrm{offset} =
(\mathrm{SFR}\,\mathrm{offset})_G -
\left\langle (\mathrm{SFR}\,\mathrm{offset})_{C_j} \right\rangle.
\end{equation}
This quantity measures the relative deviation of the reference galaxy’s star formation activity with respect to its matched controls, accounting for the intrinsic scatter within the control set itself.

Throughout this work, we interpret $\Delta \mathrm{SFR}$ offset values as follows:
\begin{itemize}[label={$\bullet$}]
    \item $\Delta \mathrm{SFR}\,\mathrm{offset} > 0$ indicates elevated star formation relative to the matched controls;
    \item $\Delta \mathrm{SFR}\,\mathrm{offset} < 0$ corresponds to suppressed star formation;
    \item $\Delta \mathrm{SFR}\,\mathrm{offset} \approx 0$ implies star formation consistent with that expected from stellar mass and environment alone.
\end{itemize}

%%%%%%%%%%%

\section{Results}
\label{sec:results}

In this section, we present a detailed analysis of how proximity to AGN influences the star formation properties of neighbouring SFGs using the EAGLE simulation. We begin by analyzing the accretion properties of the selected AGNs in our sample. We then compare the star formation activity between AGN-adjacent galaxies and a carefully matched control sample, followed by an examination of how these trends depend on stellar mass, halo mass, gas content, local density, and radial distance from the nearest AGN.

\begin{figure}[htbp!]
    \centering
    \includegraphics[width = 11cm]{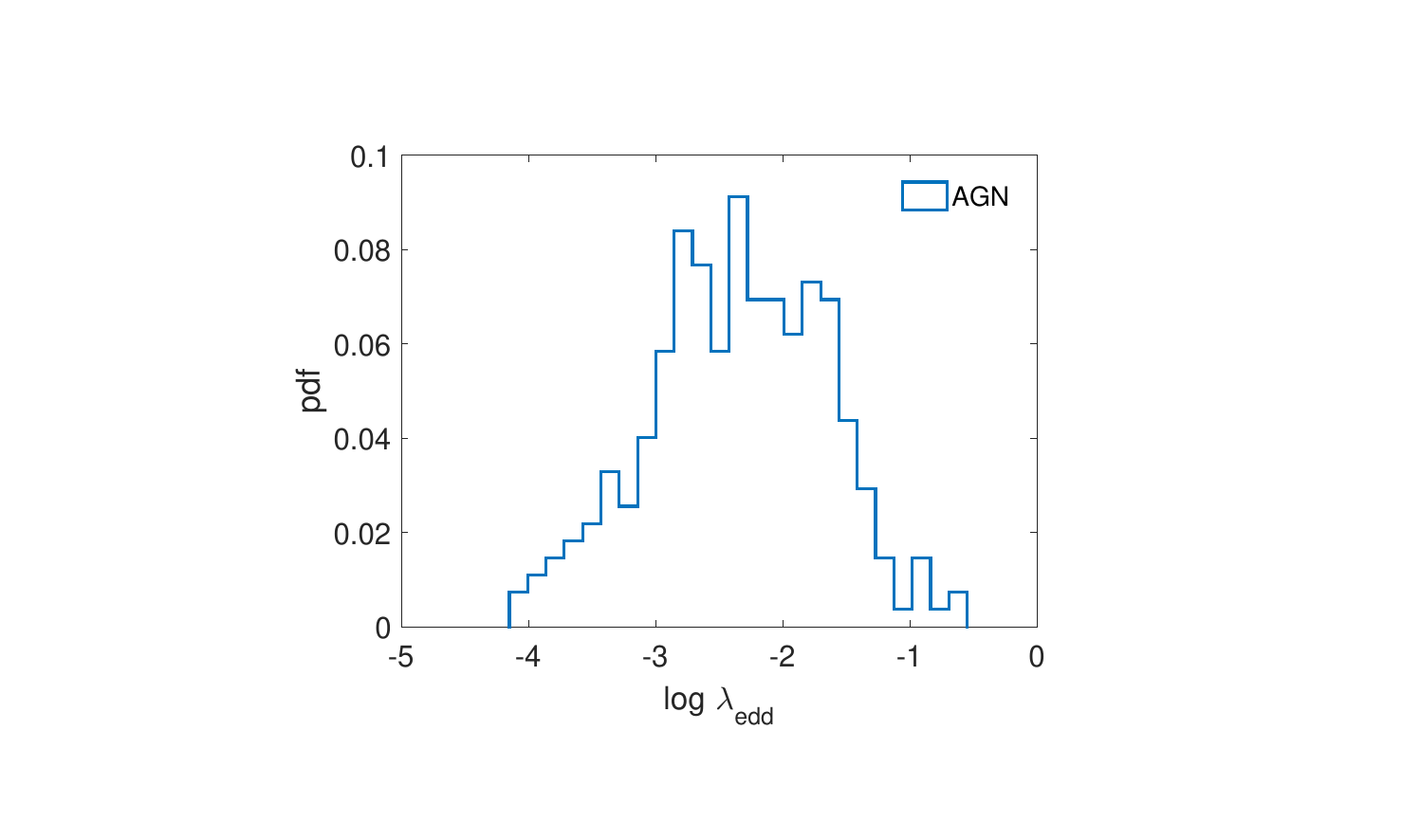}
    \includegraphics[width = 11cm]{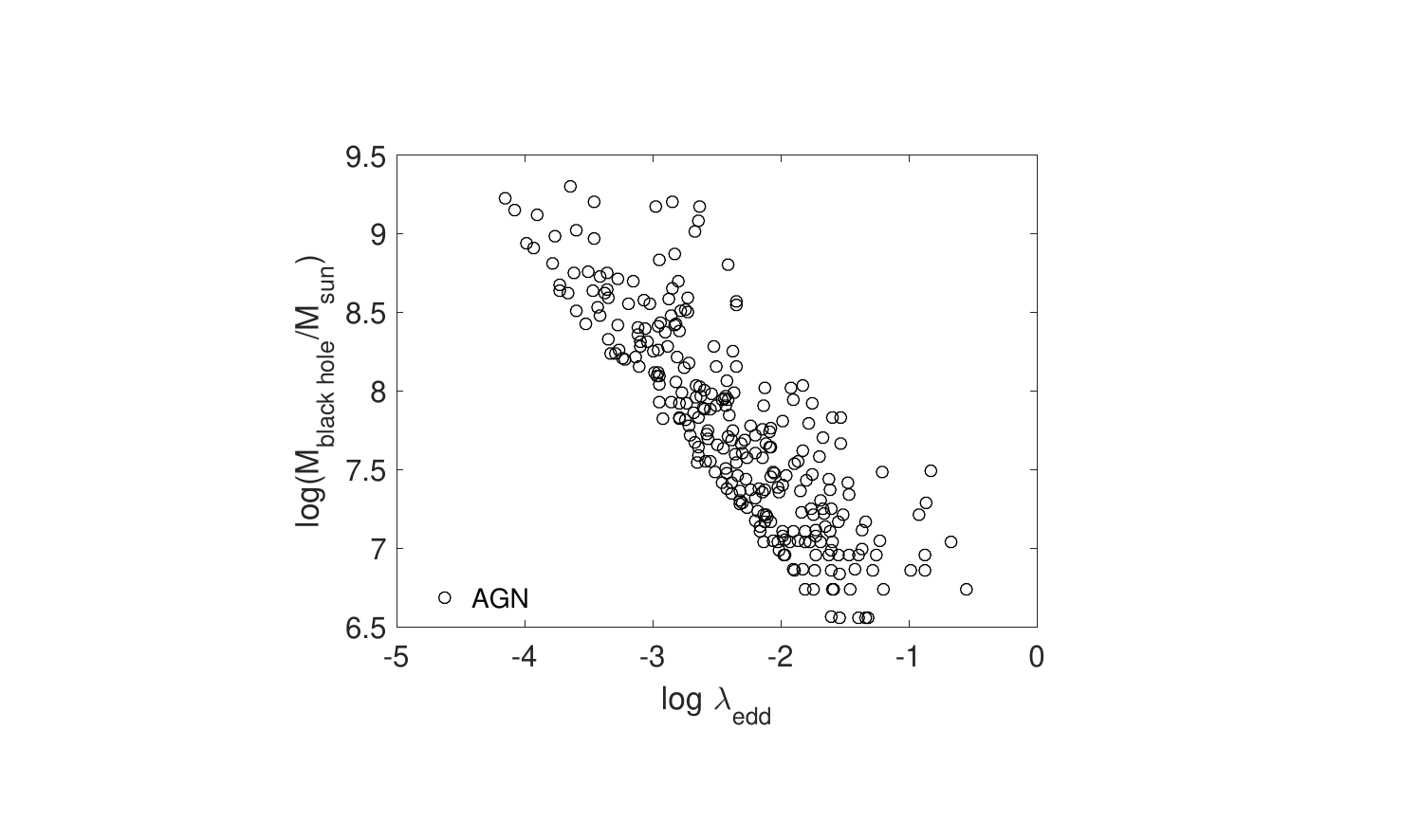}
    \caption{The top panel shows the PDF of $\log(\lambda)$, where $\lambda$ is the Eddington ratio. The distribution peaks at intermediate accretion rates ($\log(\lambda) \sim -2.5$), with relatively few AGNs accreting at very low or near-Eddington levels, indicating a preference for moderately efficient to lower accretion modes in the EAGLE simulation. The bottom panel shows a scatter plot of $\log(\lambda)$ versus black hole mass ($\log(M_{\mathrm{black\,hole}})$). An anti-correlation is visible, with the most massive black holes ($M_{\mathrm{black\,hole}} \gtrsim 10^8 M_\text{sun}$) generally accreting at lower Eddington ratios, consistent with feedback-regulated growth where massive systems evolve toward outflow-dominated regimes.}
\label{ratio}
\end{figure}

\subsection{Accretion properties of AGNs in our sample}
\label{sec:eddington}

Before examining the environmental impact of AGN activity, it is important to characterize the accretion properties of the AGNs themselves. We calculate the Eddington ratio, defined as $\lambda = \frac{L_{\mathrm{bol}}}{L_{\mathrm{edd}}}$, for each AGN in our sample. The probability distribution function of $\log(\lambda)$ is shown in top panel of \autoref{ratio}. The distribution peaks at $\log(\lambda) \sim -2.5$, indicating that the majority of AGNs accrete at relatively low Eddington ratios. Only a small fraction of objects reach near-Eddington accretion states, consistent with the EAGLE model in which AGN feedback is implemented via stochastic thermal injection, typically associated with sub-Eddington accretion conditions \citep{schaye15, crain15, guevara16}. This reinforces the picture that moderate to low accretion dominates the AGN population at $z=0$.

In bottom panel of \autoref{ratio}, we show the distribution of $\log(\lambda)$ as a function of black hole mass. We find that low-mass black holes ($M_{\mathrm{black\,hole}} \lesssim 10^{7}\,M_{\text{sun}}$) display a wide spread in accretion states, occasionally reaching the highest $\lambda$ values. In contrast, the most massive black holes ($M_{\mathrm{black\,hole}} \gtrsim 10^{9}\,M_\text{sun}$) are almost exclusively confined to very low $\lambda$, rarely exceeding $\log(\lambda) > -3$. This mass-dependent behaviour suggests that as black holes grow, their feedback becomes increasingly effective at heating and expelling gas, thereby suppressing further accretion and stabilizing them in a low-efficiency maintenance mode. Taken together, these findings highlight a dual regime: lower-mass black holes are capable of episodic high-accretion phases, while the most massive systems predominantly operate in long-term, low-accretion states.

\subsection{Star Formation Main Sequence of galaxies near AGN}
\label{sec:sfr_offsets}

To examine how the presence of nearby AGN modulates star formation in surrounding galaxies, we quantify deviations from the star formation main sequence (SFMS) using the $\Delta \mathrm{SFR}$ offset defined in Section~2.5. For each star-forming galaxy located within 2~Mpc of an AGN, the SFR offset is computed relative to a set of five AGN-isolated control galaxies matched in stellar mass and local density. This galaxy-by-galaxy comparison enables us to isolate variations in star formation activity that cannot be attributed to differences in mass or environment, and instead may be linked to AGN proximity.

\begin{figure}[h!]
\centering
\includegraphics[width = 15cm]{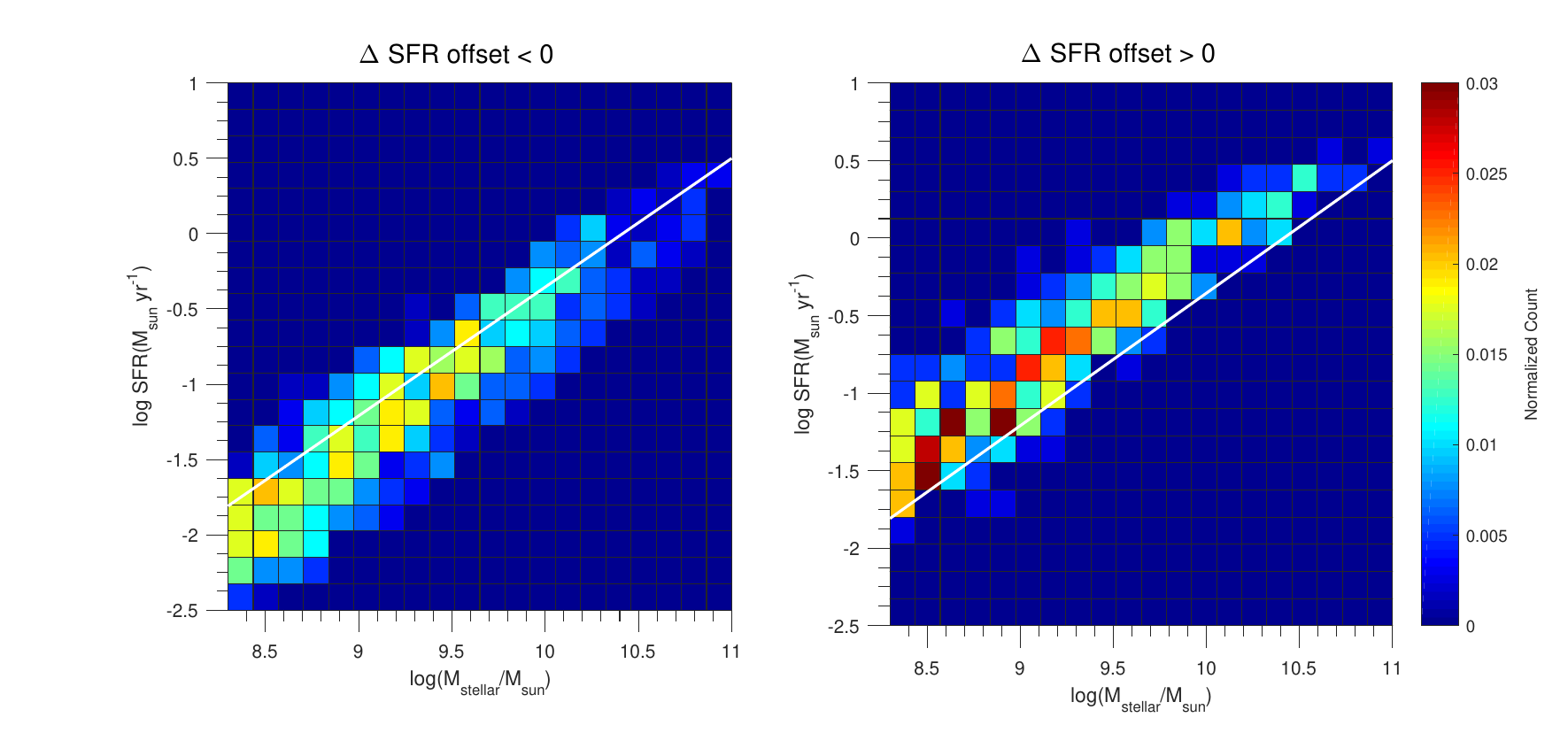}
\caption{Two-dimensional probability distributions (PDF) showing the stellar mass-SFR plane for star-forming galaxies located within 2~Mpc of an AGN. The left panel displays galaxies with suppressed star formation ($\Delta \mathrm{SFR}$ offset $<0$), while the right panel shows galaxies with enhanced star formation ($\Delta \mathrm{SFR}$ offset $>0$), where offsets are defined relative to control galaxies matched in stellar mass and local density. The colour scale indicates the normalized number density in each mass-SFR bin. The solid white line marks the best-fitting star formation main sequence (SFMS) for the combined AGN-vicinity sample. Galaxies with negative $\Delta \mathrm{SFR}$ offsets preferentially populate regions below the SFMS ridge, whereas galaxies with positive offsets are systematically shifted above it, revealing coherent and directional modulation of star formation associated with AGN proximity. The separation illustrates the differing star formation behaviour of galaxies with positive and negative $\Delta \mathrm{SFR}$ offsets. The significance of this pattern is established through comparison with the matched control sample shown in \autoref{Fig2}.
}
\label{Fig1}
\end{figure}

\begin{figure}[h!]
\centering
\includegraphics[width = 15cm]{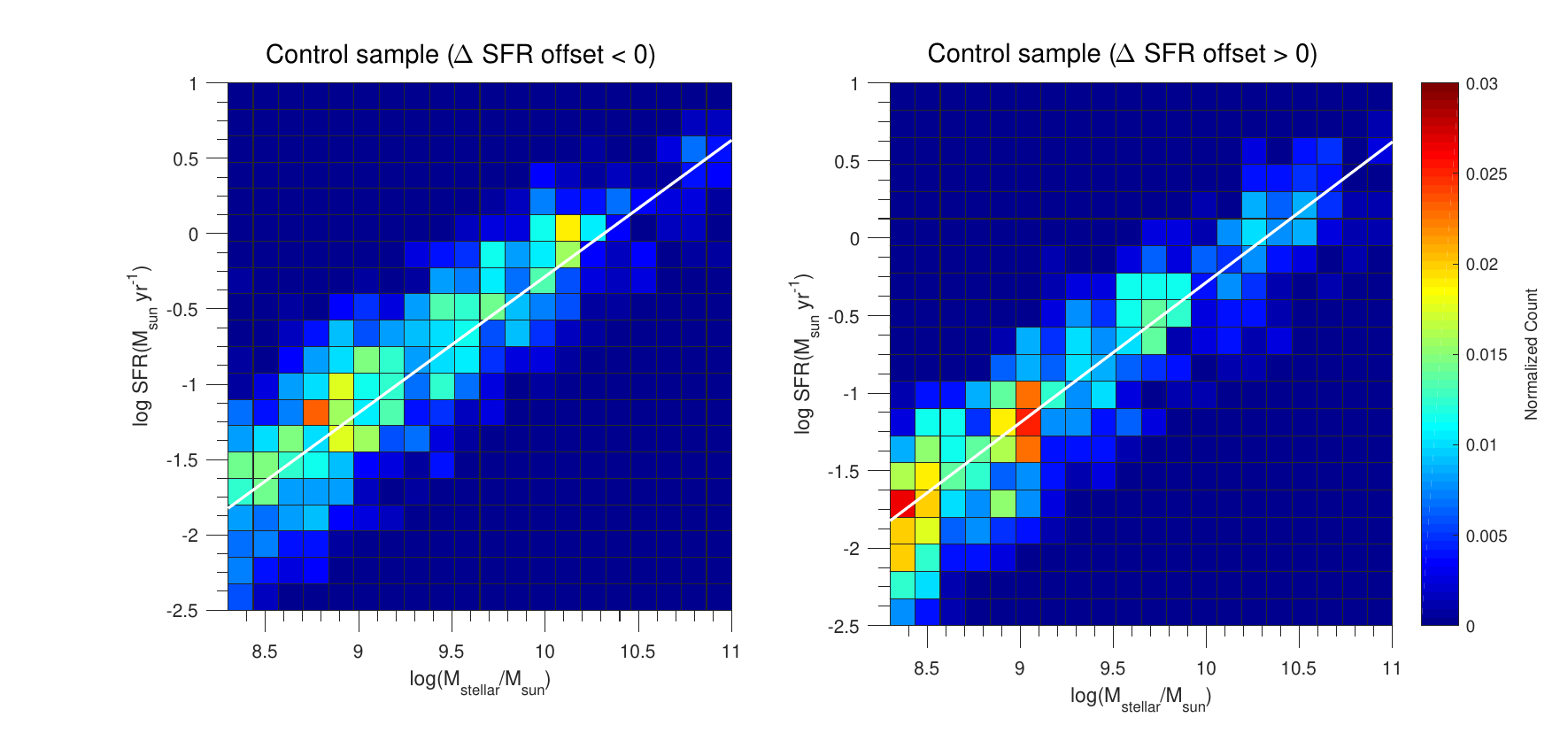}
\caption{Stellar mass-SFR distributions for the matched control galaxies (without nearby AGN) corresponding to the two AGN-vicinity subsamples shown in \autoref{Fig1}. The left and right panels present the pooled control galaxies associated with reference systems having $\Delta \mathrm{SFR}$ offset $<0$ and $\Delta \mathrm{SFR}$ offset $>0$, respectively. The solid white line denotes the best-fitting SFMS for the full control population. In contrast to the AGN-vicinity sample, both control subsets trace an indistinguishable and well-defined SFMS, with no systematic displacement above or below the ridge line. The comparison with \autoref{Fig1} demonstrates that the divergent star formation behaviour observed in AGN-adjacent galaxies is not driven by differences in stellar mass or local environment, but instead reflects a genuine non-local influence of nearby AGN activity.
}
\label{Fig2}
\end{figure}

Based on the sign of the resulting $\Delta \mathrm{SFR}$ offset, the AGN-vicinity sample is divided into two subsamples: galaxies exhibiting suppressed star formation ($\Delta \mathrm{SFR} < 0$) and those with enhanced star formation ($\Delta \mathrm{SFR} > 0$). The stellar mass--SFR distributions of these two populations are shown in \autoref{Fig1}. While both subsets broadly trace the canonical SFMS, they display systematic and opposite displacements with respect to the ridge line. Galaxies with negative $\Delta \mathrm{SFR}$ offsets preferentially populate the region below the SFMS, whereas galaxies with positive offsets are biased toward higher SFRs at fixed stellar mass, lying above the main sequence. This bifurcation demonstrates that galaxies classified according to positive and negative $\Delta \mathrm{SFR}$ offsets occupy systematically different locations relative to the star formation main sequence. However, this separation alone is expected from the definition of the $\Delta \mathrm{SFR}$ offset. The evidence that this behaviour is specifically associated with AGN proximity comes from comparison with the matched control sample (\autoref{Fig2}), where the corresponding populations do not exhibit a comparable bifurcation.

To determine whether this separation reflects a genuine influence of AGN proximity rather than the manner in which the sample is partitioned, we compare the AGN-vicinity galaxies with their matched control sample. The corresponding stellar mass-SFR relations are shown in \autoref{Fig2}. In contrast to the AGN-vicinity sample, the control galaxies associated with $\Delta \mathrm{SFR} < 0$ and $\Delta \mathrm{SFR} > 0$ reference systems both follow an indistinguishable and well-defined SFMS, with no systematic separation above or below the ridge line. This demonstrates that the divergence observed in \autoref{Fig1} is not driven by differences in stellar mass, local density, or control selection, but instead reflects a genuine modulation of star formation linked to the presence of nearby AGN.

Together, these results provide direct evidence that AGN proximity imprints a measurable and directional effect on the SFMS of neighbouring galaxies, producing both suppression and enhancement of star formation relative to otherwise similar systems in AGN-free environments.

\begin{figure}[h!]
\centering
\includegraphics[width = 11cm]{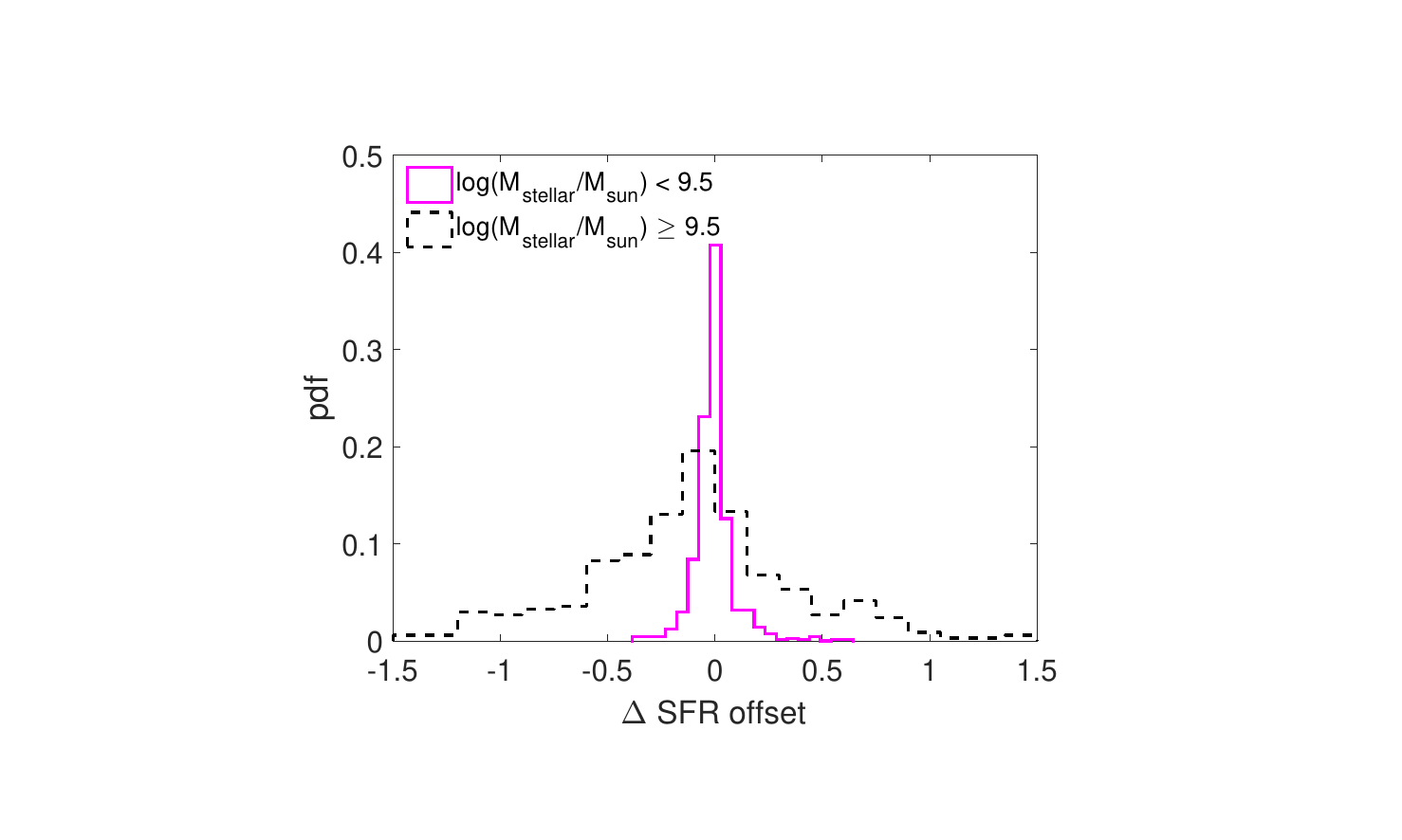}
\caption{Probability distributions of $\Delta \mathrm{SFR}$ offsets for star-forming galaxies in the AGN-vicinity sample, separated into low-mass ($M_{\mathrm{stellar}} < 10^{9.5}\,M_{\text{sun}}$) and high-mass ($M_{\mathrm{stellar}} \geq 10^{9.5}\,M_{\text{sun}}$) subsamples. Low-mass galaxies exhibit a distribution that is modestly shifted toward positive offsets but remains tightly clustered around $\Delta \mathrm{SFR} \approx 0$, indicating mild and limited enhancement of star formation in the presence of nearby AGN. In contrast, the high-mass population shows a broader distribution with an extended tail toward negative offsets, consistent with stronger and more diverse suppression of star formation relative to matched controls. The differing shapes and widths of these distributions reveal a clear mass dependence in the response of galaxies to AGN proximity, with low-mass systems displaying subtle sensitivity and massive galaxies exhibiting more pronounced regulation.}
\label{off2}
\end{figure}

\subsection{Dependence of star formation response on galaxy mass}
%\delete{Mass dependence of AGN influence} 

To investigate whether the influence of AGN proximity depends on galaxy mass, we divide the reference sample into low-mass ($M_{\mathrm{stellar}}<10^{9.5}~\mathrm{M}_{\text{sun}}$) and high-mass ($M_{\mathrm{stellar}}\geq10^{9.5}~\mathrm{M}_{\text{sun}}$) galaxies and compare their $\Delta\mathrm{SFR}$ offset distributions (\autoref{off2}). The two populations exhibit markedly different behaviours. Low-mass galaxies display a relatively narrow distribution centred close to zero, with a modest excess of positive $\Delta\mathrm{SFR}$ offsets, indicating that star formation is generally preserved or only mildly enhanced in the vicinity of AGN. In contrast, high-mass galaxies exhibit a broader distribution that is skewed towards negative offsets, implying a greater tendency for star formation suppression relative to their matched controls. A two-sample Kolmogorov--Smirnov (K--S) test confirms that the two $\Delta\mathrm{SFR}$ distributions differ significantly ($D=0.4364$, $p=4.06\times10^{-32}$), based on samples of 1,462 low-mass and 695 high-mass galaxies. These results provide strong statistical evidence that the response of neighbouring galaxies to AGN proximity depends on stellar mass, with low-mass systems being more resilient to suppression and more likely to exhibit weak star formation enhancement, while massive galaxies are more susceptible to quenching. The physical mechanisms responsible for this contrasting behaviour cannot be identified directly within the scope of the present study and therefore warrant further investigation.

\begin{figure}[htbp!]
    \centering
    \includegraphics[width = 11cm]{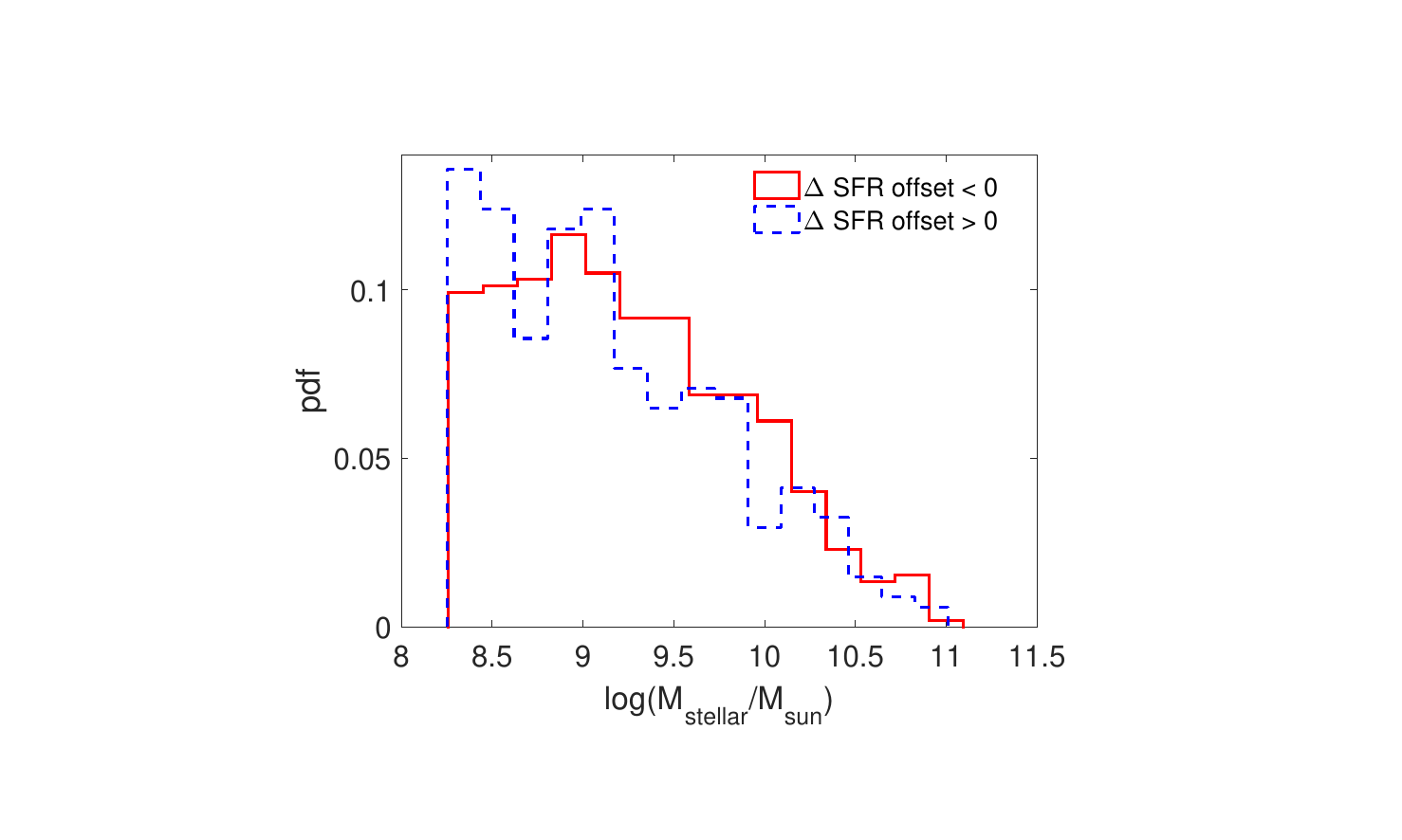}
    \includegraphics[width = 11cm]{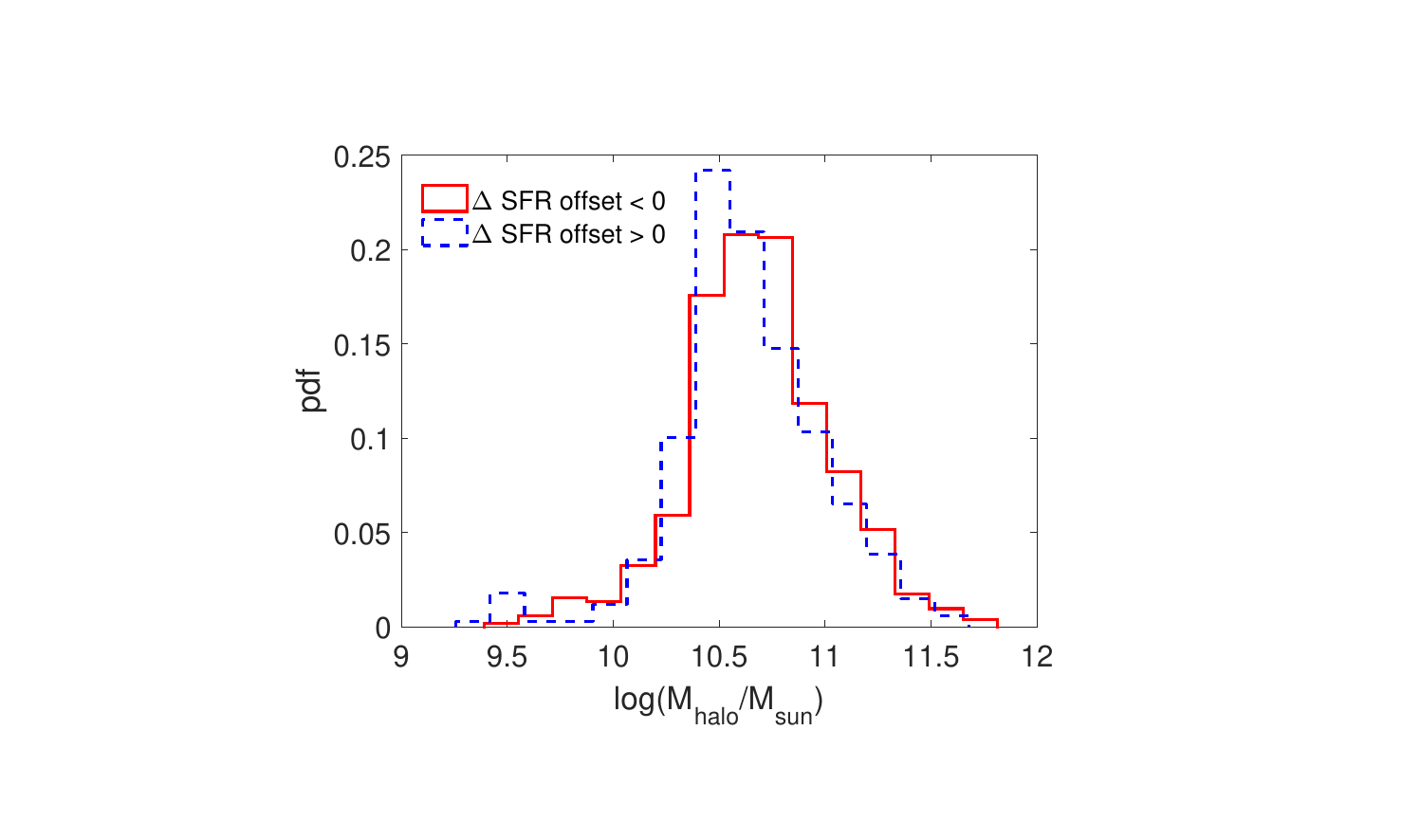}
 \caption{PDFs of stellar mass (top panel) and host dark matter halo mass (bottom panel) for star-forming galaxies located within 2~Mpc of an AGN. The samples are separated according to whether they exhibit suppressed ($\Delta \mathrm{SFR}$ offset $<0$) or enhanced ($\Delta \mathrm{SFR}$ offset $>0$) star formation relative to their matched controls. The PDFs show a mild tendency for galaxies with suppressed star formation to occupy somewhat higher stellar and halo masses than those with enhanced star formation. However, the two-sample Kolmogorov--Smirnov tests yield $D=0.0899$ ($p=0.0675$) for the stellar mass distributions and $D=0.0848$ ($p=0.0984$) for the halo mass distributions, indicating that these differences are not statistically significant. These results suggest that, although stellar and halo mass may contribute to the response of neighbouring galaxies to AGN proximity, they are unlikely to be the primary drivers of the observed star formation modulation.}
\label{Fig4}   
\end{figure}

To further examine the role of galaxy and halo mass, \autoref{Fig4} compares the stellar mass and host dark matter halo mass distributions of galaxies with suppressed ($\Delta\mathrm{SFR}$ offset $<0$) and enhanced ($\Delta\mathrm{SFR}$ offset $>0$) star formation. The probability density functions show a mild tendency for galaxies with suppressed star formation to occupy somewhat higher stellar masses and more massive host halos than those with enhanced star formation. However, the corresponding K-S tests yield $D=0.0899$ ($p=0.0675$) for the stellar mass distributions and $D=0.0848$ ($p=0.0984$) for the halo mass distributions, indicating that these differences are not statistically significant. Consequently, although stellar and halo mass appear to influence the response of neighbouring galaxies to AGN proximity, the overall mass distributions of the two populations overlap substantially. Together, these results suggest that the principal signature of mass dependence lies in the different responses of low- and high-mass galaxies to AGN proximity, rather than in a clear separation between the stellar- or halo-mass distributions of galaxies exhibiting enhanced and suppressed star formation.

\begin{figure}[htbp!]
    \centering
    \includegraphics[width = 11cm]{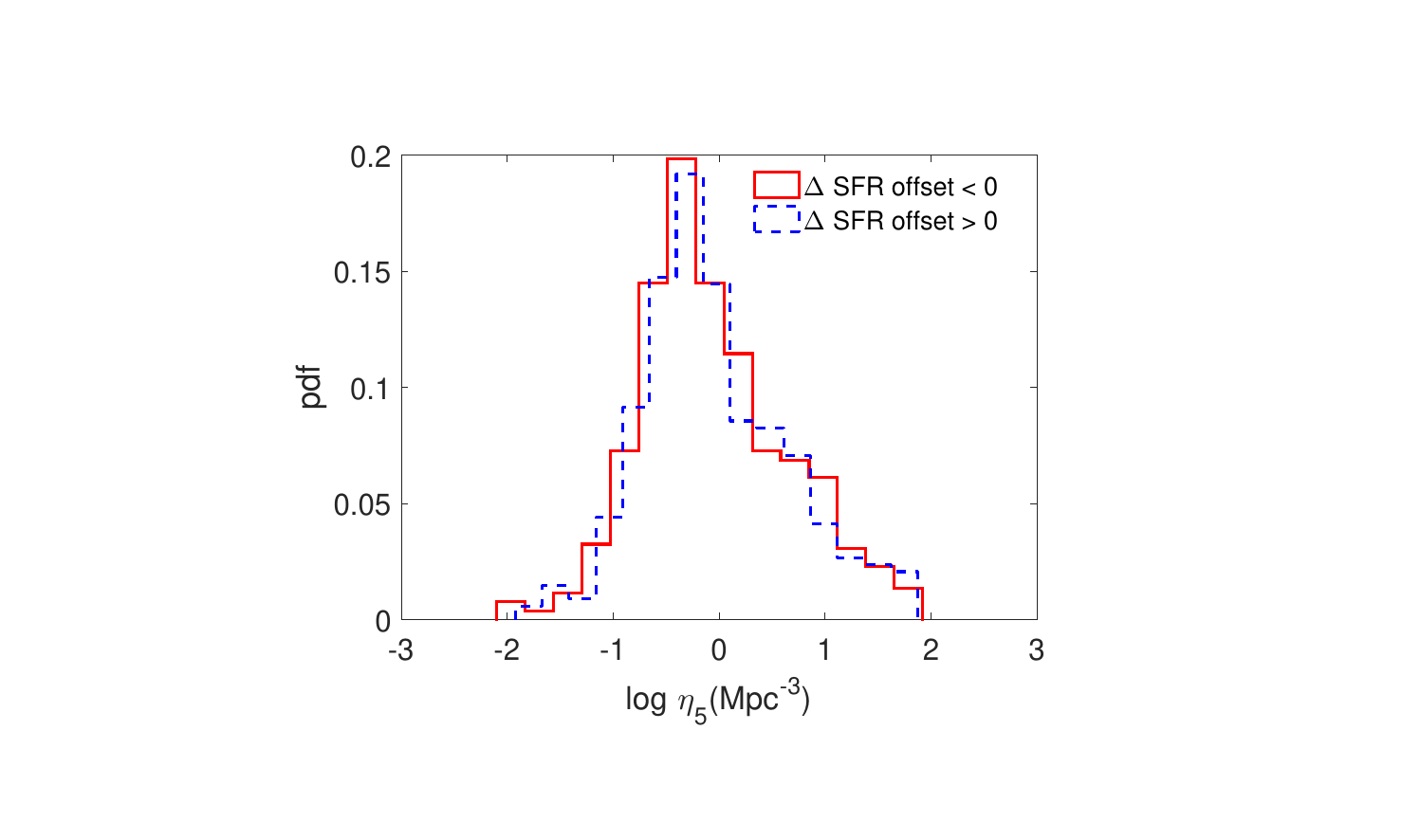}
    \caption{This compares the PDF of local environmental density of the  AGN-vicinity sample with enhanced ($\Delta \mathrm{SFR}$ offset $>0$) and suppressed ($\Delta \mathrm{SFR}$ offset $<0$) star formation. The absence of systematic differences between the two populations indicates that variations in star formation are not driven by density of their local environment but the presence of nearby AGN.}
\label{Fig5}
\end{figure}

\begin{figure}[htbp!]
    \centering
    \includegraphics[width = 11cm]{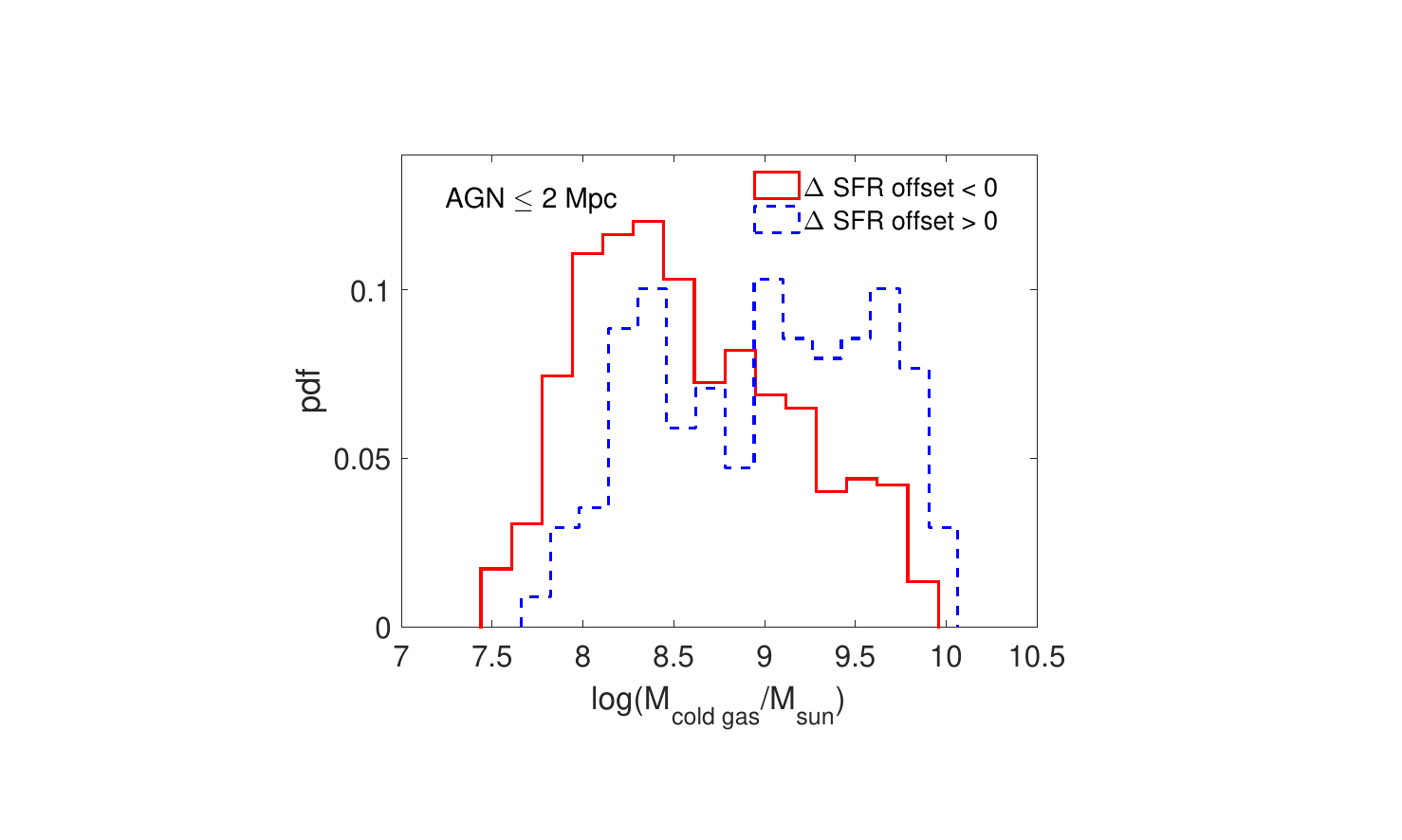}
    \includegraphics[width = 6cm]{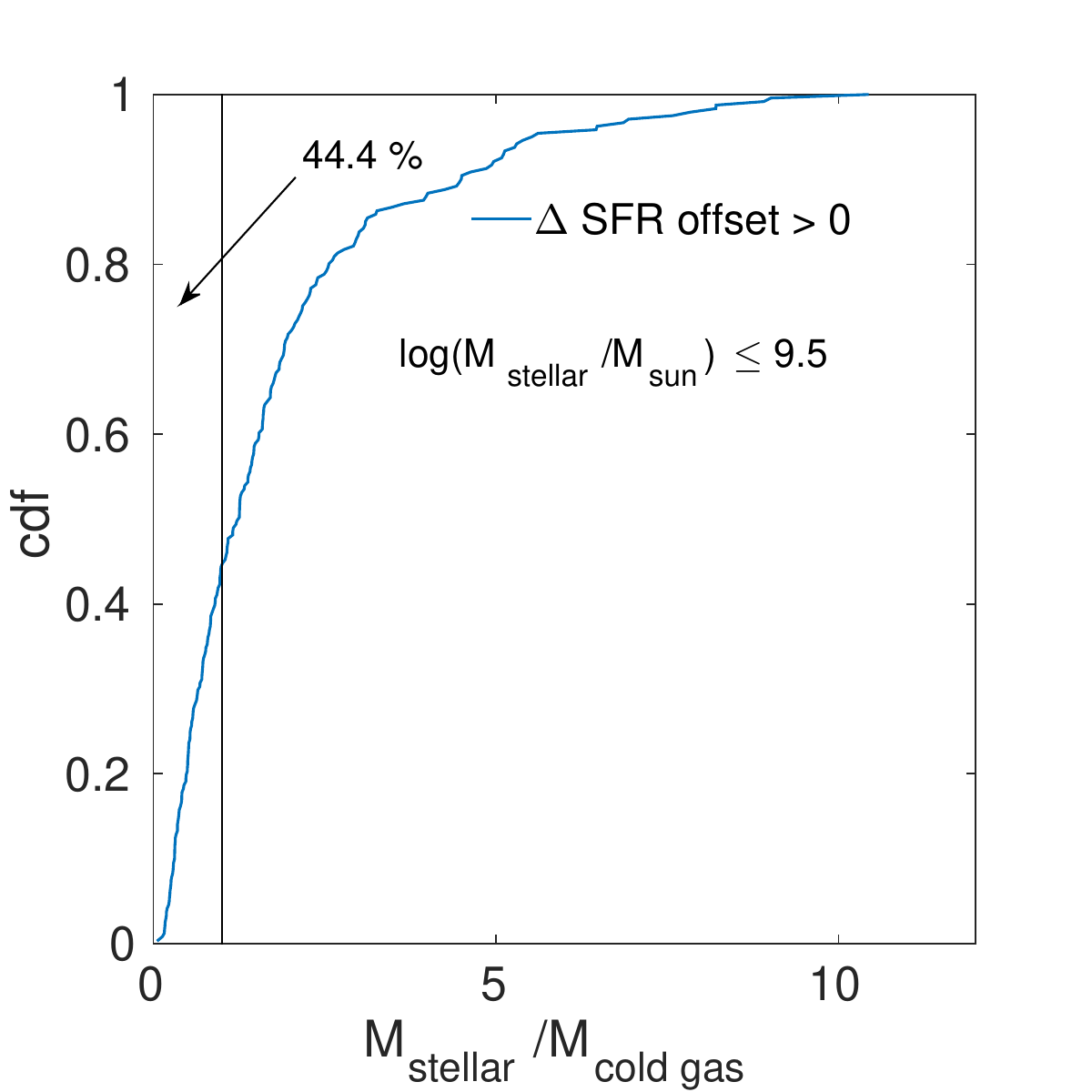} \includegraphics[width = 6cm]{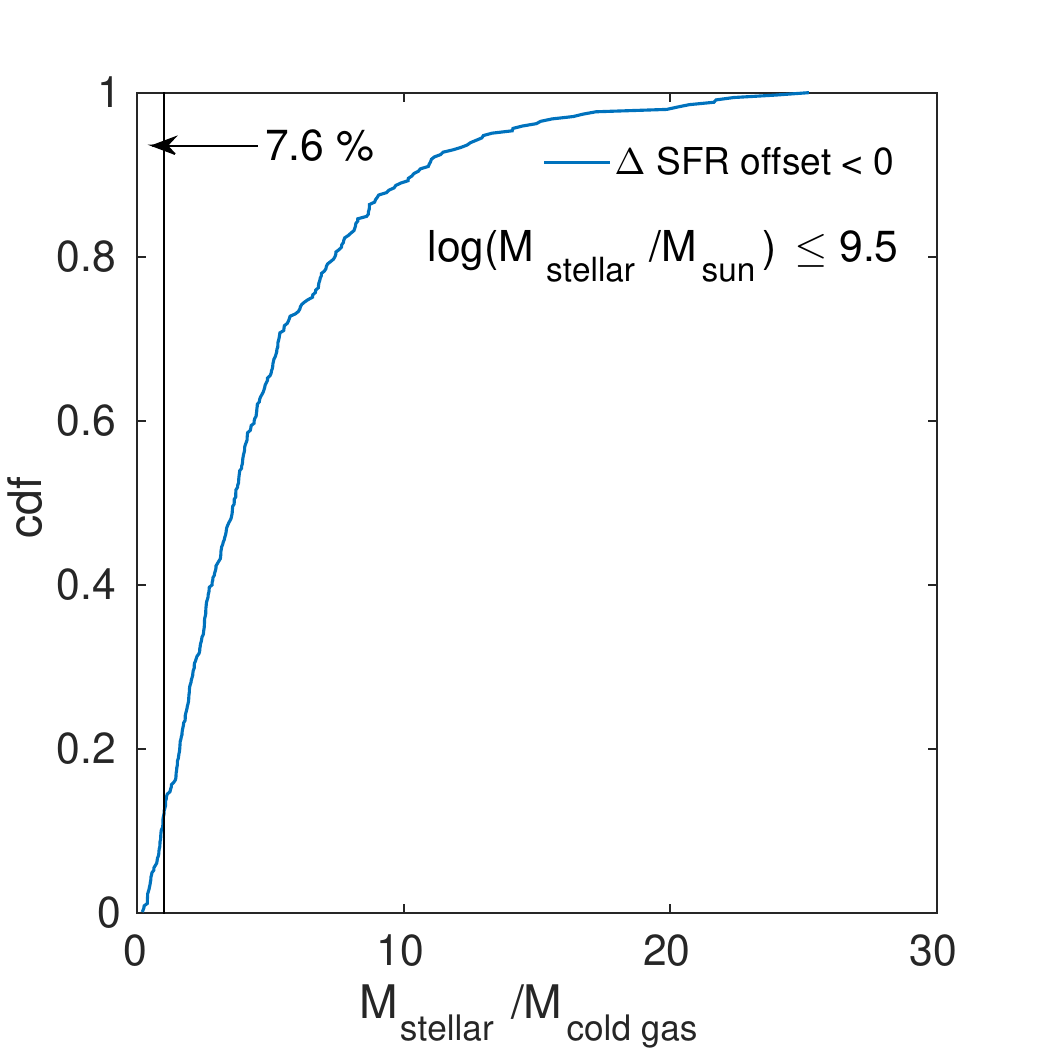} 
\caption{Top panel shows the PDFs of cold gas mass for star-forming galaxies within 2~Mpc of an AGN, separated into populations with suppressed ($\Delta \mathrm{SFR}$ offset $<0$) and enhanced ($\Delta \mathrm{SFR}$ offset $>0$) star formation. Galaxies with enhanced star formation systematically host larger cold gas reservoirs, with a pronounced excess at $M_{\mathrm{cold\,gas}} \gtrsim 10^{9}~M_{\text{sun}}$, while suppressed galaxies are comparatively gas-poor. The bottom left and bottom right panels show the CDFs of the stellar to cold gas mass ratio ($M_{\mathrm{stellar}}/M_{\mathrm{cold\,gas}}$) for low-mass ($\log M_{\mathrm{stellar}}/M_{\text{sun}} \leq 9.5$) star-forming galaxies with $\Delta \mathrm{SFR}$ offsets $>0$ and $<0$, respectively, in the AGN-vicinity sample. The vertical solid line in each panel corresponds to $M_{\mathrm{stellar}}/M_{\mathrm{cold\,gas}}=1$. While the majority of galaxies with enhanced star formation have cold gas masses comparable to or exceeding their stellar masses, only a small fraction of the star formation-suppressed galaxies exhibit such high gas fractions. This comparison demonstrates that gas-rich, low-mass galaxies are preferentially associated with star formation enhancement, whereas gas-poor systems are more likely to experience star formation suppression in the vicinity of AGN. Together, these trends indicate that cold gas availability plays a central role in mediating the response of neighbouring galaxies to AGN proximity.}
\label{Fig6}
\end{figure}

\begin{figure}[h!]
\centering
\includegraphics[width = 11cm]{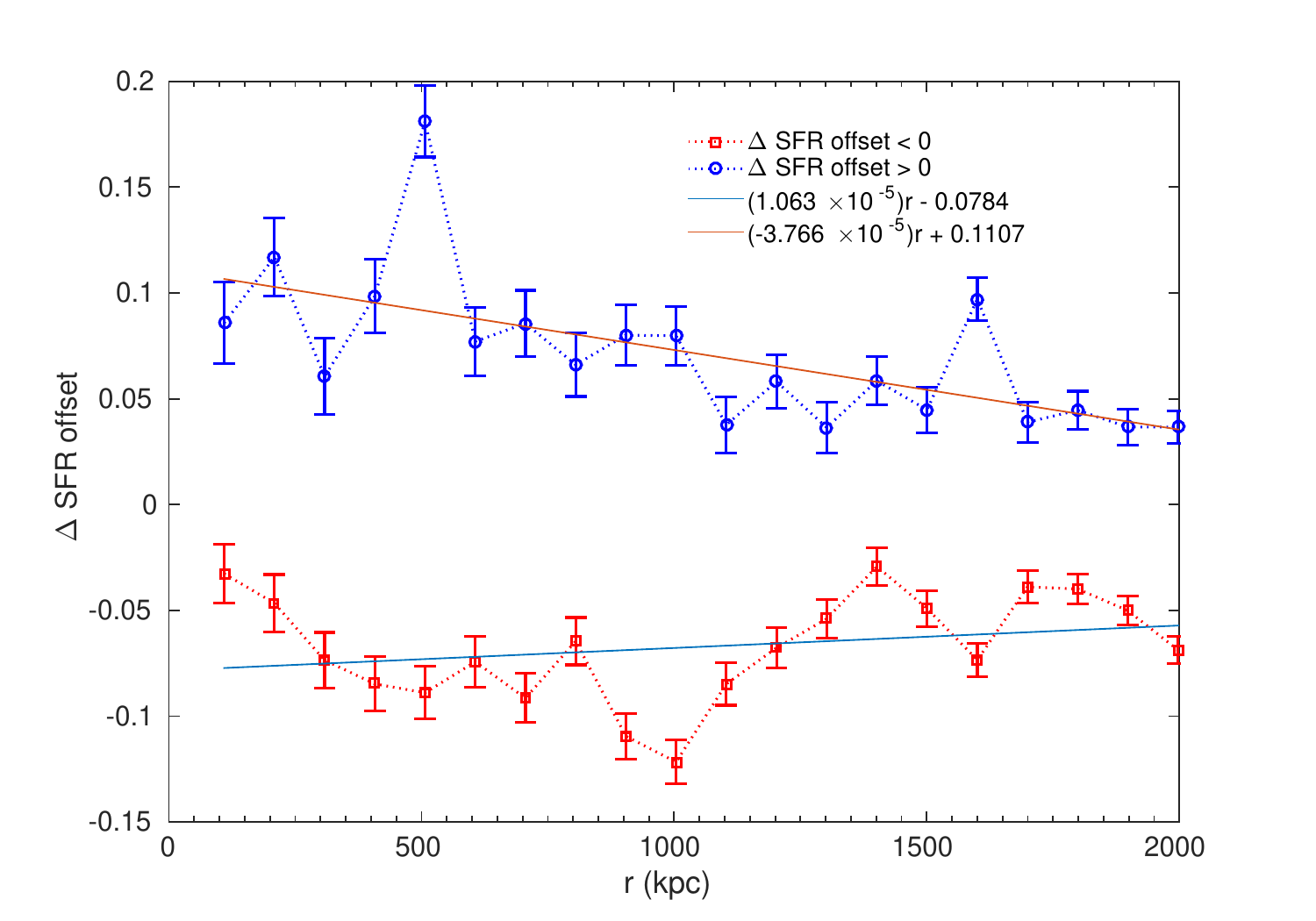}
\caption{Median $\Delta \mathrm{SFR}$ offset of star-forming galaxies as a function of physical distance from the nearest AGN, restricted to separations within 2~Mpc. The two curves correspond to galaxies with suppressed ($\Delta \mathrm{SFR}$ offset $<0$) and enhanced ($\Delta \mathrm{SFR}$ offset $>0$) star formation relative to matched control galaxies without nearby AGN. Shaded regions indicate $1\sigma$ jackknife uncertainties. The solid lines show the best-fitting linear relations for the two populations. Galaxies with suppressed star formation maintain systematically negative offsets, while enhanced systems show positive offsets at all separations, with the contrast between the two populations strongest within $\sim 1$~Mpc of the AGN. Although the magnitude of the offsets gradually decreases with distance, the persistence of non-zero values out to 2~Mpc suggests that AGN influence extends well beyond the host halo, exerting a measurable and sustained impact on the star formation activity of neighbouring galaxies.}
\label{Fig7}
\end{figure}

\subsection{Dependence on local environmental density}

Local environment is known to play an important role in regulating star formation, and for this reason local density is explicitly controlled for when constructing our reference and control samples. While the AGN-vicinity and control galaxies are statistically matched in local density by design, it remains important to examine whether star-forming galaxies exhibiting enhanced and suppressed star formation within the AGN-vicinity sample occupy systematically different environments.

To address this, we compare the local density distributions of AGN-adjacent star-forming galaxies with positive ($\Delta \mathrm{SFR} > 0$) and negative ($\Delta \mathrm{SFR} < 0$) SFR offsets, as shown in \autoref{Fig5}. The two distributions closely overlap and display no significant shift or skewness relative to one another. The two distributions closely overlap and display no significant shift or skewness relative to one another, consistent with the two-sample Kolmogorov--Smirnov (K--S) test, which yields $D=0.0459$ and $p=0.7689$. We find no evidence that galaxies with suppressed star formation preferentially reside in denser environments, nor that SFR-enhanced systems occupy systematically lower-density regions.

The absence of any statistically significant difference in local density between the two populations indicates that the divergence in star formation activity within the AGN-vicinity sample cannot be attributed to residual environmental effects. Instead, this result supports the interpretation that the observed modulation of star formation arises from processes linked to AGN proximity rather than from variations in local galaxy density. This reinforces the robustness of our control-matching strategy and strengthens the case for a non-local AGN-driven influence on star formation in neighbouring galaxies.

\subsection{Dependence on cold gas content}

Cold gas availability is a primary regulator of star formation, and any modulation of SFR in galaxies near AGN must ultimately be reflected in their gas reservoirs. To explore this connection, we compare the distributions of cold gas mass for star-forming galaxies with enhanced ($\Delta \mathrm{SFR} > 0$) and suppressed ($\Delta \mathrm{SFR} < 0$) star formation in the AGN-vicinity sample, as shown in the top panel of \autoref{Fig6}. A clear separation is evident: galaxies with enhanced star formation systematically host larger cold gas reservoirs, with a substantial fraction exceeding $M_{\mathrm{gas}} \gtrsim 10^{9}\,M_{\text{sun}}$, whereas galaxies with suppressed star formation are predominantly gas-poor. A two-sample Kolmogorov-Smirnov (K-S) test confirms that the two distributions are statistically distinct ($D=0.2838$, $p=4.45\times10^{-15}$), providing strong quantitative evidence that cold gas content differs systematically between the two populations.

While a correlation between SFR and cold gas mass is expected in any star-forming population, the contrast observed here is not a trivial manifestation of that relation. Instead, the excess of gas-rich systems among galaxies with positive $\Delta \mathrm{SFR}$ offsets suggests that proximity to AGN may actively modulate how existing gas reservoirs are utilized. In particular, AGN-driven outflows or shocks may compress cold gas in neighbouring galaxies, temporarily enhancing star formation rather than suppressing it.

The role of gas content becomes even clearer when combined with the stellar-mass dependence discussed earlier. Although the stellar-mass distributions of galaxies with positive and negative $\Delta \mathrm{SFR}$ offsets show only a mild, statistically insignificant difference (\autoref{Fig4}), galaxies with positive $\Delta \mathrm{SFR}$ offsets exhibit significantly larger cold gas reservoirs, indicating that cold gas availability is more strongly associated with the observed modulation of star formation than stellar mass alone. To quantify this behaviour, we examine the stellar to cold gas mass ratio for low-mass galaxies with positive and negative $\Delta \mathrm{SFR}$ offsets, shown in the bottom left and bottom right panels of \autoref{Fig6}, respectively. Nearly $44\%$ of the galaxies with positive $\Delta \mathrm{SFR}$ offsets have stellar to gas mass ratios below unity, and approximately $75\%$ have ratios below two, indicating that they are overwhelmingly gas rich. In contrast, only $\sim7.6\%$ of the galaxies with negative $\Delta \mathrm{SFR}$ offsets have stellar to gas mass ratios below unity. This direct comparison demonstrates that the high gas fractions are not simply a generic property of low-mass galaxies, but are preferentially associated with galaxies exhibiting enhanced star formation in the vicinity of AGN.

Such high gas fractions are consistent with observational studies of low-mass galaxies \citep{catinella18} and with predictions from hydrodynamical simulations \citep{dave17}. Our results therefore suggest that gas richness is a more important factor than stellar mass alone in determining how galaxies respond to nearby AGN. The direct comparison between the star formation-enhanced and star formation-suppressed low-mass populations further demonstrates that high gas fractions are preferentially associated with galaxies exhibiting enhanced star formation, rather than being a generic characteristic of low-mass galaxies. Gas-rich, low-mass systems appear particularly susceptible to star formation triggering, whereas gas-poor galaxies are more prone to suppression. This highlights cold gas content as a key mediator of the non-local influence of AGN on star formation.

\subsection{Radial dependence of star formation suppression and enhancement}

We next examine how the modulation of star formation in AGN-adjacent galaxies depends on their physical distance from the nearest AGN. \autoref{Fig7} shows the median $\Delta \mathrm{SFR}$ offset as a function of AGN-galaxy separation for star-forming galaxies within 2\,Mpc. The two curves correspond to galaxies classified as star formation-suppressed ($\Delta \mathrm{SFR} < 0$) and star formation-enhanced ($\Delta \mathrm{SFR} > 0$) relative to their matched control samples.

A clear radial structure emerges. Galaxies with suppressed star formation exhibit systematically negative median $\Delta \mathrm{SFR}$ offsets at all separations, while those with enhanced star formation maintain positive offsets across the full radial range considered. To quantify the radial behaviour, we fitted linear relations to the median $\Delta \mathrm{SFR}$ offsets as a function of AGN-galaxy separation. The suppressed population is described by $\Delta \mathrm{SFR}=(1.063\times10^{-5})\,r-0.0784$, while the enhanced population follows $\Delta \mathrm{SFR}=(-3.766\times10^{-5})\,r+0.1107$, where $r$ is the AGN-galaxy separation in kpc. These fits provide a quantitative characterization of the gradual radial trends, with the magnitude of both star formation suppression and enhancement decreasing with increasing AGN--galaxy separation. Although the linear model does not fully capture the scatter in the data, the fitted linear relations provide a quantitative characterization of the gradual radial trends exhibited by the two populations. The opposite slopes of the two fits indicate that the magnitude of star formation suppression gradually decreases with increasing distance from the AGN, while the enhancement similarly weakens with separation. These quantitative trends are consistent with the largest departures from the matched control samples occurring within approximately $1$\,Mpc of the AGN and becoming progressively weaker at larger separations. However, the median offsets remain significantly offset from zero even at the largest distances probed.

The persistence of non-zero median $\Delta \mathrm{SFR}$ offsets out to 2\,Mpc suggests that the influence associated with AGN proximity is not confined to the host halo but extends into the broader galactic environment. Although the fitted gradients are relatively shallow, the opposite radial trends for the suppressed and enhanced populations demonstrate that the modulation of star formation is not distance-independent. Instead, the gradual convergence of the two populations with increasing separation supports the interpretation that AGN-related processes operate over extended spatial scales while coupling differently to neighbouring galaxies according to their internal properties, such as their stellar mass and cold gas content. These results therefore strengthen the case that the observed modulation of star formation reflects a genuine physical dependence on AGN proximity rather than a systematic effect unrelated to distance.

\section{Conclusions}
\label{sec:conclusion}

In this study, we have examined the influence of active galactic nuclei (AGN) on the star formation activity of neighbouring galaxies using the EAGLE cosmological hydrodynamical simulation. By focusing on star-forming galaxies located within 2~Mpc of AGN hosts and comparing them with carefully constructed control samples matched in stellar mass and local density, we isolate the role of AGN proximity from intrinsic galaxy properties and environmental effects. This approach enables a direct and statistically robust assessment of how AGN feedback manifests beyond the confines of the host galaxy.

Our analysis demonstrates that AGN proximity leaves a measurable, yet non-uniform, imprint on the star formation properties of surrounding galaxies. The distribution of $\Delta \mathrm{SFR}$ offsets reveals both modest suppression and enhancement relative to matched controls, indicating that AGN can act as regulators of star formation in multiple modes. Low- and high-mass galaxies exhibit systematically different responses to AGN proximity, with low-mass, gas-rich systems more likely to show star formation enhancement and high-mass, gas-poor systems tending towards suppression. Although the stellar-mass distributions of the enhanced and suppressed populations are not themselves statistically distinct, the contrasting responses indicate that cold gas availability, together with galaxy mass, plays an important role in determining how neighbouring galaxies respond to AGN-driven environmental influence.

A key result of this work is that these variations cannot be attributed to differences in local galaxy density. Within the AGN-vicinity sample, galaxies exhibiting enhanced and suppressed star formation display statistically indistinguishable density distributions, ruling out local overdensity as the dominant driver of the observed $\Delta \mathrm{SFR}$ behaviour. Instead, the modulation of star formation appears intrinsically linked to the presence of nearby AGN.

We further identify a clear spatial dependence in this effect. Quantitative linear fits show that the contrast between star formation enhancement and suppression is strongest within $\sim1$~Mpc of AGN hosts, weakens gradually with increasing separation, and remains detectable out to 2~Mpc. This persistence highlights the ability of AGN feedback to propagate through the circumgalactic and intergalactic medium, influencing neighbouring galaxies well beyond the host halo. Such non-local effects underscore the role of AGN as environmental agents rather than purely internal regulators.

Taken together, our results demonstrate that AGN feedback operates across a broad range of spatial scales and gives rise to a bifurcated response in neighbouring galaxies. The interplay between AGN-driven energy injection, cold gas availability, and galaxy mass produces a complex landscape in which both suppression and triggering of star formation can occur. This environmental dimension of AGN feedback adds an important layer to our understanding of galaxy evolution, linking small-scale accretion physics to large-scale structure.

The main conclusions of this work can be summarized as follows:
\begin{enumerate}[label=(\roman*)]
    \item Proximity to AGN leads to statistically significant modulation of star formation in neighbouring galaxies, producing both enhancement and suppression relative to mass- and density-matched control samples.
    \item The response to AGN proximity is strongly dependent on galaxy properties. Low-mass, gas-rich systems are more likely to exhibit star formation enhancement, whereas massive, gas-poor galaxies tend towards suppression. The direct comparison of low-mass galaxies with enhanced and suppressed star formation further demonstrates that high gas fractions are preferentially associated with the enhanced population, highlighting cold gas availability as a key factor governing the response of neighbouring galaxies to nearby AGN.
    \item Differences in star formation activity within the AGN-vicinity sample are not driven by variations in local density, indicating that AGN proximity is the primary differentiating factor.
    \item The influence of AGN extends to distances of at least 2~Mpc, with the strongest effects occurring within $\sim 1$~Mpc, revealing AGN as non-local regulators of star formation.
\end{enumerate}

These findings build upon and extend previous results obtained with the EAGLE simulation. Despite employing a single-mode, stochastic thermal prescription for AGN feedback \citep{schaye15}, EAGLE reproduces a wide range of observed galaxy properties, suggesting that this simplified approach captures the essential physics of AGN-driven heating. Our results show that even within this framework, AGN feedback can generate subtle yet persistent signatures in the surrounding galaxy population.

Previous studies have demonstrated that AGN activity can shorten quenching timescales in massive satellites by heating or expelling gas \citep{wright19}, and that sustained AGN heating combined with halted gas accretion can exhaust the interstellar medium \citep{visser25}. Our work complements these findings by showing that AGN influence is not confined to the host or satellite systems alone, but can extend outward to affect star formation in neighbouring galaxies on megaparsec scales. Similarly, while black hole mass and central potential depth are known to be strong predictors of quenching in central galaxies \citep{bluck23}, our results highlight an additional environmental channel through which AGN can impact external systems.

Our conclusions are also consistent with recent work by \citet{goubert24}, who find that quiescence in massive galaxies is primarily governed by intrinsic properties, whereas environmental processes dominate for low-mass systems. We extend this picture by demonstrating that proximity to AGN can further modulate star formation in gas-rich, low-mass galaxies, potentially accelerating or initiating quenching. This interpretation aligns with the results of \citet{dashyan19}, who show that AGN feedback from central galaxies can suppress star formation in satellites out to several virial radii. Together, these studies reinforce the view that AGN feedback operates as a non-local, environmentally mediated process.

A caveat of our analysis is the adopted lower stellar mass limit of $\log(M_{\mathrm{stellar}}/M_{\text{sun}}) \geq 8.3$, for which the lowest-mass galaxies are resolved by $\sim 110$ baryonic particles. While increasing this threshold improves numerical fidelity, it substantially reduces the available sample size and limits statistical power. Our chosen mass cut therefore represents a balance between resolution and robustness, and future higher-resolution simulations will be essential to further refine these results.

In conclusion, this work provides new evidence that AGN feedback exerts a spatially extended and environmentally dependent influence on galaxy evolution. By revealing distinct pathways for star formation suppression and enhancement in neighbouring galaxies, our study highlights the importance of non-local AGN feedback in shaping galaxy populations. Upcoming facilities such as \textit{JWST} \citep{jwst06}, SKA \citep{ska09}, and \textit{Euclid} \citep{euclid25}, together with next-generation simulations, will be crucial for testing these predictions and elucidating the physical mechanisms that govern AGN-driven environmental regulation.

\section{Acknowledgements}
AD is supported by a postdoctoral fellowship from Harish-Chandra Research Institute, Allahabad. B.P. gratefully acknowledges the Anusandhan National Research Foundation (ANRF), Government of India, for financial support under project number ANRF/ARG/2025/000535/PS. BP also acknowledges the support provided by Inter-University Centre for Astronomy and Astrophysics (IUCAA), Pune, through an associateship programme. The authors extend their gratitude to the Virgo Consortium for providing public access to their simulation data. The EAGLE simulations were performed using the DiRAC-2 facility at Durham, managed by the ICC, and the PRACE facility Curie based in France at TGCC, CEA, Bruy\`{e}res-le-Ch\^{a}tel.

\section{Data availability}
The EAGLE simulation data analysed in this work are publicly accessible via the EAGLE database at \url{https://icc.dur.ac.uk/Eagle/database.php}
. Any additional data products generated as part of this study can be made available by the authors upon reasonable request.

\bibliographystyle{raa}
\bibliography{agn}

@ARTICLE{fabian99,
       author = {{Fabian}, Andrew C.},
        title = "{Active Galactic Nuclei}",
      journal = {Proceedings of the National Academy of Science},
         year = 1999,
        month = apr,
       volume = {96},
       number = {9},
        pages = {4749-4751},
          doi = {10.1073/pnas.96.9.4749},
       adsurl = {https://ui.adsabs.harvard.edu/abs/1999PNAS...96.4749F}
}

@ARTICLE{woo02,
       author = {{Woo}, Jong-Hak and {Urry}, C. Megan},
        title = "{Active Galactic Nucleus Black Hole Masses and Bolometric Luminosities}",
      journal = {\apj},
         year = 2002,
        month = nov,
       volume = {579},
       number = {2},
        pages = {530-544},
          doi = {10.1086/342878},
archivePrefix = {arXiv},
       eprint = {astro-ph/0207249},
 primaryClass = {astro-ph},
       adsurl = {https://ui.adsabs.harvard.edu/abs/2002ApJ...579..530W}
}

@ARTICLE{kawata05,
       author = {{Kawata}, D. and {Gibson}, B.~K.},
        title = "{Self-regulated active galactic nuclei heating in elliptical galaxies}",
      journal = {\mnras},
         year = 2005,
        month = mar,
       volume = {358},
       number = {1},
        pages = {L16-L20},
          doi = {10.1111/j.1745-3933.2005.00018.x},
archivePrefix = {arXiv},
       eprint = {astro-ph/0409068},
 primaryClass = {astro-ph},
       adsurl = {https://ui.adsabs.harvard.edu/abs/2005MNRAS.358L..16K}
}

@ARTICLE{wagner13,
       author = {{Wagner}, A.~Y. and {Umemura}, M. and {Bicknell}, G.~V.},
        title = "{Ultrafast Outflows: Galaxy-scale Active Galactic Nucleus Feedback}",
      journal = {\apjl},
         year = 2013,
        month = jan,
       volume = {763},
       number = {1},
          eid = {L18},
        pages = {L18},
          doi = {10.1088/2041-8205/763/1/L18},
archivePrefix = {arXiv},
       eprint = {1211.5851},
 primaryClass = {astro-ph.CO},
       adsurl = {https://ui.adsabs.harvard.edu/abs/2013ApJ...763L..18W}
}

@ARTICLE{morganti17,
       author = {{Morganti}, Raffaella},
        title = "{The many routes to AGN feedback}",
      journal = {Frontiers in Astronomy and Space Sciences},
         year = 2017,
        month = nov,
       volume = {4},
          eid = {42},
        pages = {42},
          doi = {10.3389/fspas.2017.00042},
archivePrefix = {arXiv},
       eprint = {1712.05301},
 primaryClass = {astro-ph.GA},
       adsurl = {https://ui.adsabs.harvard.edu/abs/2017FrASS...4...42M}
}

@ARTICLE{baron18,
       author = {{Baron}, Dalya and {Netzer}, Hagai and {Prochaska}, J. Xavier and {Cai}, Zheng and {Cantalupo}, Sebastiano and {Martin}, D. Christopher and {Matuszewski}, Mateusz and {Moore}, Anna M. and {Morrissey}, Patrick and {Neill}, James D.},
        title = "{Direct evidence of AGN feedback: a post-starburst galaxy stripped of its gas by AGN-driven winds}",
      journal = {\mnras},
         year = 2018,
        month = nov,
       volume = {480},
       number = {3},
        pages = {3993-4016},
          doi = {10.1093/mnras/sty2113},
archivePrefix = {arXiv},
       eprint = {1804.03150},
 primaryClass = {astro-ph.GA},
       adsurl = {https://ui.adsabs.harvard.edu/abs/2018MNRAS.480.3993B}
}

@ARTICLE{santoro20,
       author = {{Santoro}, F. and {Tadhunter}, C. and {Baron}, D. and {Morganti}, R. and {Holt}, J.},
        title = "{AGN-driven outflows and the AGN feedback efficiency in young radio galaxies}",
      journal = {\aap},
         year = 2020,
        month = dec,
       volume = {644},
          eid = {A54},
        pages = {A54},
          doi = {10.1051/0004-6361/202039077},
archivePrefix = {arXiv},
       eprint = {2009.11175},
 primaryClass = {astro-ph.GA},
       adsurl = {https://ui.adsabs.harvard.edu/abs/2020A&A...644A..54S}
}

@ARTICLE{somerville08,
       author = {{Somerville}, Rachel S. and {Hopkins}, Philip F. and {Cox}, Thomas J. and {Robertson}, Brant E. and {Hernquist}, Lars},
        title = "{A semi-analytic model for the co-evolution of galaxies, black holes and active galactic nuclei}",
      journal = {\mnras},
         year = 2008,
        month = dec,
       volume = {391},
       number = {2},
        pages = {481-506},
          doi = {10.1111/j.1365-2966.2008.13805.x},
archivePrefix = {arXiv},
       eprint = {0808.1227},
 primaryClass = {astro-ph},
       adsurl = {https://ui.adsabs.harvard.edu/abs/2008MNRAS.391..481S}
}

@ARTICLE{kormendy13,
       author = {{Kormendy}, John and {Ho}, Luis C.},
        title = "{Coevolution (Or Not) of Supermassive Black Holes and Host Galaxies}",
      journal = {\araa},
         year = 2013,
        month = aug,
       volume = {51},
       number = {1},
        pages = {511-653},
          doi = {10.1146/annurev-astro-082708-101811},
archivePrefix = {arXiv},
       eprint = {1304.7762},
 primaryClass = {astro-ph.CO},
       adsurl = {https://ui.adsabs.harvard.edu/abs/2013ARA&A..51..511K}
}

@ARTICLE{heckman14,
       author = {{Heckman}, Timothy M. and {Best}, Philip N.},
        title = "{The Coevolution of Galaxies and Supermassive Black Holes: Insights from Surveys of the Contemporary Universe}",
      journal = {\araa},
         year = 2014,
        month = aug,
       volume = {52},
        pages = {589-660},
          doi = {10.1146/annurev-astro-081913-035722},
archivePrefix = {arXiv},
       eprint = {1403.4620},
 primaryClass = {astro-ph.GA},
       adsurl = {https://ui.adsabs.harvard.edu/abs/2014ARA&A..52..589H}
}

@ARTICLE{harrison17,
       author = {{Harrison}, C.~M.},
        title = "{Impact of supermassive black hole growth on star formation}",
      journal = {Nature Astronomy},
         year = 2017,
        month = jul,
       volume = {1},
          eid = {0165},
        pages = {0165},
          doi = {10.1038/s41550-017-0165},
archivePrefix = {arXiv},
       eprint = {1703.06889},
 primaryClass = {astro-ph.GA},
       adsurl = {https://ui.adsabs.harvard.edu/abs/2017NatAs...1E.165H}
}

@ARTICLE{springel05,
       author = {{Springel}, Volker and {Di Matteo}, Tiziana and {Hernquist}, Lars},
        title = "{Modelling feedback from stars and black holes in galaxy mergers}",
      journal = {\mnras},
         year = 2005,
        month = aug,
       volume = {361},
       number = {3},
        pages = {776-794},
          doi = {10.1111/j.1365-2966.2005.09238.x},
archivePrefix = {arXiv},
       eprint = {astro-ph/0411108},
 primaryClass = {astro-ph},
       adsurl = {https://ui.adsabs.harvard.edu/abs/2005MNRAS.361..776S}
}

@ARTICLE{baldry04,
       author = {{Baldry}, Ivan K. and {Glazebrook}, Karl and {Brinkmann}, Jon and {Ivezi{\'c}}, {\v{Z}}eljko and {Lupton}, Robert H. and {Nichol}, Robert C. and {Szalay}, Alexander S.},
        title = "{Quantifying the Bimodal Color-Magnitude Distribution of Galaxies}",
      journal = {\apj},
         year = 2004,
        month = jan,
       volume = {600},
       number = {2},
        pages = {681-694},
          doi = {10.1086/380092},
archivePrefix = {arXiv},
       eprint = {astro-ph/0309710},
 primaryClass = {astro-ph},
       adsurl = {https://ui.adsabs.harvard.edu/abs/2004ApJ...600..681B}
}

@ARTICLE{das25,
       author = {{Das}, Apashanka and {Pandey}, Biswajit},
        title = "{The long road to the Green Valley: Tracing the evolution of the Green Valley galaxies in the EAGLE simulation}",
      journal = {\jcap},
         year = 2025,
        month = may,
       volume = {2025},
       number = {5},
          eid = {101},
        pages = {101},
          doi = {10.1088/1475-7516/2025/05/101},
archivePrefix = {arXiv},
       eprint = {2501.01207},
 primaryClass = {astro-ph.GA},
       adsurl = {https://ui.adsabs.harvard.edu/abs/2025JCAP...05..101D}
}

@ARTICLE{ruffa19,
       author = {{Ruffa}, Ilaria and {Davis}, Timothy A. and {Prandoni}, Isabella and {Laing}, Robert A. and {Paladino}, Rosita and {Parma}, Paola and {de Ruiter}, Hans and {Casasola}, Viviana and {Bureau}, Martin and {Warren}, Joshua},
        title = "{The AGN fuelling/feedback cycle in nearby radio galaxies - II. Kinematics of the molecular gas}",
      journal = {\mnras},
         year = 2019,
        month = nov,
       volume = {489},
       number = {3},
        pages = {3739-3757},
          doi = {10.1093/mnras/stz2368},
archivePrefix = {arXiv},
       eprint = {1908.09229},
 primaryClass = {astro-ph.GA},
       adsurl = {https://ui.adsabs.harvard.edu/abs/2019MNRAS.489.3739R}
}

@ARTICLE{shangguan20,
       author = {{Shangguan}, Jinyi and {Ho}, Luis C. and {Bauer}, Franz E. and {Wang}, Ran and {Treister}, Ezequiel},
        title = "{AGN Feedback and Star Formation of Quasar Host Galaxies: Insights from the Molecular Gas}",
      journal = {\apj},
         year = 2020,
        month = aug,
       volume = {899},
       number = {2},
          eid = {112},
        pages = {112},
          doi = {10.3847/1538-4357/aba8a1},
archivePrefix = {arXiv},
       eprint = {2007.11286},
 primaryClass = {astro-ph.GA},
       adsurl = {https://ui.adsabs.harvard.edu/abs/2020ApJ...899..112S}
}

@ARTICLE{ellison21,
       author = {{Ellison}, Sara L. and {Wong}, Tony and {S{\'a}nchez}, Sebastian F. and {Colombo}, Dario and {Bolatto}, Alberto and {Barrera-Ballesteros}, Jorge and {Garc{\'\i}a-Benito}, Rub{\'e}n and {Kalinova}, Veselina and {Luo}, Yufeng and {Rubio}, Monica and {Vogel}, Stuart N.},
        title = "{The EDGE-CALIFA survey: central molecular gas depletion in AGN host galaxies - a smoking gun for quenching?}",
      journal = {\mnras},
         year = 2021,
        month = jul,
       volume = {505},
       number = {1},
        pages = {L46-L51},
          doi = {10.1093/mnrasl/slab047},
archivePrefix = {arXiv},
       eprint = {2105.02916},
 primaryClass = {astro-ph.GA},
       adsurl = {https://ui.adsabs.harvard.edu/abs/2021MNRAS.505L..46E}
}

@ARTICLE{sampaio23,
       author = {{Sampaio}, V.~M. and {Arag{\'o}n-Salamanca}, A. and {Merrifield}, M.~R. and {de Carvalho}, R.~R. and {Zhou}, S. and {Ferreras}, I.},
        title = "{The co-evolution of strong AGN and central galaxies in different environments}",
      journal = {\mnras},
         year = 2023,
        month = oct,
       volume = {524},
       number = {4},
        pages = {5327-5339},
          doi = {10.1093/mnras/stad2211},
archivePrefix = {arXiv},
       eprint = {2307.10435},
 primaryClass = {astro-ph.GA},
       adsurl = {https://ui.adsabs.harvard.edu/abs/2023MNRAS.524.5327S}
}

@ARTICLE{aird21,
       author = {{Aird}, James and {Coil}, Alison L.},
        title = "{The AGN-galaxy-halo connection: the distribution of AGN host halo masses to z = 2.5}",
      journal = {\mnras},
         year = 2021,
        month = apr,
       volume = {502},
       number = {4},
        pages = {5962-5980},
          doi = {10.1093/mnras/stab312},
archivePrefix = {arXiv},
       eprint = {2010.02957},
 primaryClass = {astro-ph.GA},
       adsurl = {https://ui.adsabs.harvard.edu/abs/2021MNRAS.502.5962A}
}

@ARTICLE{lopes17,
       author = {{Lopes}, P.~A.~A. and {Ribeiro}, A.~L.~B. and {Rembold}, S.~B.},
        title = "{NoSOCS in SDSS - VI. The environmental dependence of AGN in clusters and  field in the local Universe}",
      journal = {\mnras},
         year = 2017,
        month = nov,
       volume = {472},
       number = {1},
        pages = {409-418},
          doi = {10.1093/mnras/stx2046},
archivePrefix = {arXiv},
       eprint = {1708.02242},
 primaryClass = {astro-ph.GA},
       adsurl = {https://ui.adsabs.harvard.edu/abs/2017MNRAS.472..409L}
}

@ARTICLE{ceccarelli21,
       author = {{Ceccarelli}, Laura and {Duplancic}, Fernanda and {Garcia Lambas}, Diego},
        title = "{The impact of void environment on AGN}",
      journal = {\mnras},
         year = 2021,
        month = nov,
       volume = {509},
       number = {2},
        pages = {1805-1819},
          doi = {10.1093/mnras/stab2902},          
archivePrefix = {arXiv},
       eprint = {2111.11488},
 primaryClass = {astro-ph.GA},
       adsurl = {https://ui.adsabs.harvard.edu/abs/2021arXiv211111488C}
}

@ARTICLE{banerjee25,
       author = {{Banerjee}, Amrita and {Pandey}, Biswajit and {Nandi}, Anindita},
        title = "{Clustering and physical properties of AGN and Star-Forming Galaxies at fixed stellar mass: does assembly bias have a role in AGN activity?}",
      journal = {\pasa},
     year = 2025,
        month = jun,
       volume = {42},
       number = {78},
        pages = {1-16},
          doi = {10.1017/pasa.2025.10052},
archivePrefix = {arXiv},
       eprint = {2310.12943},
 primaryClass = {astro-ph.GA},
       adsurl = {https://ui.adsabs.harvard.edu/abs/2023arXiv231012943B}
}

@ARTICLE{gilli09,
       author = {{Gilli}, R. and {Zamorani}, G. and {Miyaji}, T. and {Silverman}, J. and {Brusa}, M. and {Mainieri}, V. and {Cappelluti}, N. and {Daddi}, E. and {Porciani}, C. and {Pozzetti}, L. and {Civano}, F. and {Comastri}, A. and {Finoguenov}, A. and {Fiore}, F. and {Salvato}, M. and {Vignali}, C. and {Hasinger}, G. and {Lilly}, S. and {Impey}, C. and {Trump}, J. and {Capak}, P. and {McCracken}, H. and {Scoville}, N. and {Taniguchi}, Y. and {Carollo}, C.~M. and {Contini}, T. and {Kneib}, J. -P. and {Le Fevre}, O. and {Renzini}, A. and {Scodeggio}, M. and {Bardelli}, S. and {Bolzonella}, M. and {Bongiorno}, A. and {Caputi}, K. and {Cimatti}, A. and {Coppa}, G. and {Cucciati}, O. and {de La Torre}, S. and {de Ravel}, L. and {Franzetti}, P. and {Garilli}, B. and {Iovino}, A. and {Kampczyk}, P. and {Knobel}, C. and {Kova{\v{c}}}, K. and {Lamareille}, F. and {Le Borgne}, J. -F. and {Le Brun}, V. and {Maier}, C. and {Mignoli}, M. and {Pell{\`o}}, R. and {Peng}, Y. and {Perez Montero}, E. and {Ricciardelli}, E. and {Tanaka}, M. and {Tasca}, L. and {Tresse}, L. and {Vergani}, D. and {Zucca}, E. and {Abbas}, U. and {Bottini}, D. and {Cappi}, A. and {Cassata}, P. and {Fumana}, M. and {Guzzo}, L. and {Leauthaud}, A. and {Maccagni}, D. and {Marinoni}, C. and {Memeo}, P. and {Meneux}, B. and {Oesch}, P. and {Scaramella}, R. and {Walcher}, J.},
        title = "{The spatial clustering of X-ray selected AGN in the XMM-COSMOS field}",
      journal = {\aap},
         year = 2009,
        month = jan,
       volume = {494},
       number = {1},
        pages = {33-48},
          doi = {10.1051/0004-6361:200810821},
archivePrefix = {arXiv},
       eprint = {0810.4769},
 primaryClass = {astro-ph},
       adsurl = {https://ui.adsabs.harvard.edu/abs/2009A&A...494...33G}
}

@ARTICLE{zhang21,
       author = {{Zhang}, Ziwen and {Wang}, Huiyuan and {Luo}, Wentao and {Mo}, H.~J. and {Liang}, Zhixiong and {Li}, Ran and {Yang}, Xiaohu and {Wang}, Tinggui and {Zhang}, Hongxin and {Hong}, Hui and {Wang}, Xiaoyu and {Wang}, Enci and {Li}, Pengfei and {Shi}, JingJing},
        title = "{Hosts and triggers of AGNs in the Local Universe}",
      journal = {\aap},
         year = 2021,
        month = jun,
       volume = {650},
          eid = {A155},
        pages = {A155},
          doi = {10.1051/0004-6361/202040150},
archivePrefix = {arXiv},
       eprint = {2012.10640},
 primaryClass = {astro-ph.GA},
       adsurl = {https://ui.adsabs.harvard.edu/abs/2021A&A...650A.155Z}
}

@ARTICLE{singh23,
       author = {{Singh}, Ankit and {Park}, Changbom and {Choi}, Ena and {Kim}, Juhan and {Jun}, Hyunsung and {Gibson}, Brad K. and {Kim}, Yonghwi and {Lee}, Jaehyun and {Snaith}, Owain},
        title = "{On the Effects of Local Environment on Active Galactic Nucleus (AGN) in the Horizon Run 5 Simulation}",
      journal = {\apj},
         year = 2023,
        month = aug,
       volume = {953},
       number = {1},
          eid = {64},
        pages = {64},
          doi = {10.3847/1538-4357/acdd6b},
archivePrefix = {arXiv},
       eprint = {2308.01584},
 primaryClass = {astro-ph.GA},
       adsurl = {https://ui.adsabs.harvard.edu/abs/2023ApJ...953...64S}
}

@ARTICLE{satyapal14,
       author = {{Satyapal}, Shobita and {Ellison}, Sara L. and {McAlpine}, William and {Hickox}, Ryan C. and {Patton}, David R. and {Mendel}, J. Trevor},
        title = "{Galaxy pairs in the Sloan Digital Sky Survey - IX. Merger-induced AGN activity as traced by the Wide-field Infrared Survey Explorer}",
      journal = {\mnras},
         year = 2014,
        month = jun,
       volume = {441},
       number = {2},
        pages = {1297-1304},
          doi = {10.1093/mnras/stu650},
archivePrefix = {arXiv},
       eprint = {1403.7531},
 primaryClass = {astro-ph.GA},
       adsurl = {https://ui.adsabs.harvard.edu/abs/2014MNRAS.441.1297S}
}

@ARTICLE{mandelbaum09,
       author = {{Mandelbaum}, Rachel and {Li}, Cheng and {Kauffmann}, Guinevere and {White}, Simon D.~M.},
        title = "{Halo masses for optically selected and for radio-loud AGN from clustering and galaxy-galaxy lensing}",
      journal = {\mnras},
         year = 2009,
        month = feb,
       volume = {393},
       number = {2},
        pages = {377-392},
          doi = {10.1111/j.1365-2966.2008.14235.x},
archivePrefix = {arXiv},
       eprint = {0806.4089},
 primaryClass = {astro-ph},
       adsurl = {https://ui.adsabs.harvard.edu/abs/2009MNRAS.393..377M}
}

@ARTICLE{donoso14,
       author = {{Donoso}, E. and {Yan}, Lin and {Stern}, D. and {Assef}, R.~J.},
        title = "{The Angular Clustering of WISE-selected Active Galactic Nuclei: Different Halos for Obscured and Unobscured Active Galactic Nuclei}",
      journal = {\apj},
         year = 2014,
        month = jul,
       volume = {789},
       number = {1},
          eid = {44},
        pages = {44},
          doi = {10.1088/0004-637X/789/1/44},
archivePrefix = {arXiv},
       eprint = {1309.2277},
 primaryClass = {astro-ph.CO},
       adsurl = {https://ui.adsabs.harvard.edu/abs/2014ApJ...789...44D}
}

@ARTICLE{hale18,
       author = {{Hale}, C.~L. and {Jarvis}, M.~J. and {Delvecchio}, I. and {Hatfield}, P.~W. and {Novak}, M. and {Smol{\v{c}}i{\'c}}, V. and {Zamorani}, G.},
        title = "{The clustering and bias of radio-selected AGN and star-forming galaxies in the COSMOS field}",
      journal = {\mnras},
         year = 2018,
        month = mar,
       volume = {474},
       number = {3},
        pages = {4133-4150},
          doi = {10.1093/mnras/stx2954},
archivePrefix = {arXiv},
       eprint = {1711.05201},
 primaryClass = {astro-ph.GA},
       adsurl = {https://ui.adsabs.harvard.edu/abs/2018MNRAS.474.4133H}
}

@ARTICLE{oei23,
       author = {{Oei}, Martijn S.~S.~L. and {van Weeren}, Reinout J. and {Gast}, Aivin R.~D.~J.~G.~I.~B. and {Botteon}, Andrea and {Hardcastle}, Martin J. and {Dabhade}, Pratik and {Shimwell}, Tim W. and {R{\"o}ttgering}, Huub J.~A. and {Drabent}, Alexander},
        title = "{Measuring the giant radio galaxy length distribution with the LoTSS}",
      journal = {\aap},
         year = 2023,
        month = apr,
       volume = {672},
          eid = {A163},
        pages = {A163},
          doi = {10.1051/0004-6361/202243572},
archivePrefix = {arXiv},
       eprint = {2210.10234},
 primaryClass = {astro-ph.GA},
       adsurl = {https://ui.adsabs.harvard.edu/abs/2023A&A...672A.163O}
}

@ARTICLE{hardcastle19,
       author = {{Hardcastle}, M.~J. and {Williams}, W.~L. and {Best}, P.~N. and {Croston}, J.~H. and {Duncan}, K.~J. and {R{\"o}ttgering}, H.~J.~A. and {Sabater}, J. and {Shimwell}, T.~W. and {Tasse}, C. and {Callingham}, J.~R. and {Cochrane}, R.~K. and {de Gasperin}, F. and {G{\"u}rkan}, G. and {Jarvis}, M.~J. and {Mahatma}, V. and {Miley}, G.~K. and {Mingo}, B. and {Mooney}, S. and {Morabito}, L.~K. and {O'Sullivan}, S.~P. and {Prandoni}, I. and {Shulevski}, A. and {Smith}, D.~J.~B.},
        title = "{Radio-loud AGN in the first LoTSS data release. The lifetimes and environmental impact of jet-driven sources}",
      journal = {\aap},
         year = 2019,
        month = feb,
       volume = {622},
          eid = {A12},
        pages = {A12},
          doi = {10.1051/0004-6361/201833893},
archivePrefix = {arXiv},
       eprint = {1811.07943},
 primaryClass = {astro-ph.GA},
       adsurl = {https://ui.adsabs.harvard.edu/abs/2019A&A...622A..12H}
}

@ARTICLE{dhabade20,
       author = {{Dabhade}, P. and {R{\"o}ttgering}, H.~J.~A. and {Bagchi}, J. and {Shimwell}, T.~W. and {Hardcastle}, M.~J. and {Sankhyayan}, S. and {Morganti}, R. and {Jamrozy}, M. and {Shulevski}, A. and {Duncan}, K.~J.},
        title = "{Giant radio galaxies in the LOFAR Two-metre Sky Survey. I. Radio and environmental properties}",
      journal = {\aap},
         year = 2020,
        month = mar,
       volume = {635},
          eid = {A5},
        pages = {A5},
          doi = {10.1051/0004-6361/201935589},
archivePrefix = {arXiv},
       eprint = {1904.00409},
 primaryClass = {astro-ph.GA},
       adsurl = {https://ui.adsabs.harvard.edu/abs/2020A&A...635A...5D}
}

@ARTICLE{oei22,
       author = {{Oei}, Martijn S.~S.~L. and {van Weeren}, Reinout J. and {Hardcastle}, Martin J. and {Botteon}, Andrea and {Shimwell}, Tim W. and {Dabhade}, Pratik and {Gast}, Aivin R.~D.~J.~G.~I.~B. and {R{\"o}ttgering}, Huub J.~A. and {Br{\"u}ggen}, Marcus and {Tasse}, Cyril and {Williams}, Wendy L. and {Shulevski}, Aleksandar},
        title = "{The discovery of a radio galaxy of at least 5 Mpc}",
      journal = {\aap},
         year = 2022,
        month = apr,
       volume = {660},
          eid = {A2},
        pages = {A2},
          doi = {10.1051/0004-6361/202142778},
archivePrefix = {arXiv},
       eprint = {2202.05427},
 primaryClass = {astro-ph.GA},
       adsurl = {https://ui.adsabs.harvard.edu/abs/2022A&A...660A...2O}
}

@ARTICLE{schaye15,
       author = {{Schaye}, Joop and {Crain}, Robert A. and {Bower}, Richard G. and {Furlong}, Michelle and {Schaller}, Matthieu and {Theuns}, Tom and {Dalla Vecchia}, Claudio and {Frenk}, Carlos S. and {McCarthy}, I.~G. and {Helly}, John C. and {Jenkins}, Adrian and {Rosas-Guevara}, Y.~M. and {White}, Simon D.~M. and {Baes}, Maarten and {Booth}, C.~M. and {Camps}, Peter and {Navarro}, Julio F. and {Qu}, Yan and {Rahmati}, Alireza and {Sawala}, Till and {Thomas}, Peter A. and {Trayford}, James},
        title = "{The EAGLE project: simulating the evolution and assembly of galaxies and their environments}",
      journal = {\mnras},
         year = 2015,
        month = jan,
       volume = {446},
       number = {1},
        pages = {521-554},
          doi = {10.1093/mnras/stu2058},
archivePrefix = {arXiv},
       eprint = {1407.7040},
 primaryClass = {astro-ph.GA},
       adsurl = {https://ui.adsabs.harvard.edu/abs/2015MNRAS.446..521S}
}

@ARTICLE{dashyan19,
       author = {{Dashyan}, Gohar and {Choi}, Ena and {Somerville}, Rachel S. and {Naab}, Thorsten and {Quirk}, Amanda C.~N. and {Hirschmann}, Michaela and {Ostriker}, Jeremiah P.},
        title = "{AGN-driven quenching of satellite galaxies}",
      journal = {\mnras},
         year = 2019,
        month = aug,
       volume = {487},
       number = {4},
        pages = {5889-5901},
          doi = {10.1093/mnras/stz1697},
archivePrefix = {arXiv},
       eprint = {1906.07431},
 primaryClass = {astro-ph.GA},
       adsurl = {https://ui.adsabs.harvard.edu/abs/2019MNRAS.487.5889D}
}

@ARTICLE{wright19,
       author = {{Wright}, Ruby J. and {Lagos}, Claudia del P. and {Davies}, Luke J.~M. and {Power}, Chris and {Trayford}, James W. and {Wong}, O. Ivy},
        title = "{Quenching time-scales of galaxies in the EAGLE simulations}",
      journal = {\mnras},
         year = 2019,
        month = aug,
       volume = {487},
       number = {3},
        pages = {3740-3758},
          doi = {10.1093/mnras/stz1410},
archivePrefix = {arXiv},
       eprint = {1810.07335},
 primaryClass = {astro-ph.GA},
       adsurl = {https://ui.adsabs.harvard.edu/abs/2019MNRAS.487.3740W}
}

@ARTICLE{goubert24,
       author = {{Goubert}, Paul H. and {Bluck}, Asa F.~L. and {Piotrowska}, Joanna M. and {Maiolino}, Roberto},
        title = "{The role of environment and AGN feedback in quenching local galaxies: comparing cosmological hydrodynamical simulations to the SDSS}",
      journal = {\mnras},
         year = 2024,
        month = mar,
       volume = {528},
       number = {3},
        pages = {4891-4921},
          doi = {10.1093/mnras/stae269},
archivePrefix = {arXiv},
       eprint = {2401.12953},
 primaryClass = {astro-ph.GA},
       adsurl = {https://ui.adsabs.harvard.edu/abs/2024MNRAS.528.4891G}
}

@ARTICLE{bluck23,
       author = {{Bluck}, Asa F.~L. and {Piotrowska}, Joanna M. and {Maiolino}, Roberto},
        title = "{The Fundamental Signature of Star Formation Quenching from AGN Feedback: A Critical Dependence of Quiescence on Supermassive Black Hole Mass, Not Accretion Rate}",
      journal = {\apj},
         year = 2023,
        month = feb,
       volume = {944},
       number = {1},
          eid = {108},
        pages = {108},
          doi = {10.3847/1538-4357/acac7c},
archivePrefix = {arXiv},
       eprint = {2301.03677},
 primaryClass = {astro-ph.GA},
       adsurl = {https://ui.adsabs.harvard.edu/abs/2023ApJ...944..108B}
}

@ARTICLE{visser25,
       author = {{Visser-Zadvornyi}, Anatolii I. and {Carstairs}, Mary E. and {Oman}, Kyle A. and {Verheijen}, Marc A.~W.},
        title = "{Star formation and stellar \& AGN feedback in the absence of accretion, not gas stripping, set the quenching time-scale in satellite galaxies}",
      journal = {\mnras},
         year = 2025,
        month = jun,
       volume = {540},
       number = {2},
        pages = {1730-1744},
          doi = {10.1093/mnras/staf802},
archivePrefix = {arXiv},
       eprint = {2503.15183},
 primaryClass = {astro-ph.GA},
       adsurl = {https://ui.adsabs.harvard.edu/abs/2025MNRAS.540.1730V}
}

@ARTICLE{catinella18,
       author = {{Catinella}, Barbara and {Saintonge}, Am{\'e}lie and {Janowiecki}, Steven and {Cortese}, Luca and {Dav{\'e}}, Romeel and {Lemonias}, Jenna J. and {Cooper}, Andrew P. and {Schiminovich}, David and {Hummels}, Cameron B. and {Fabello}, Silvia and {Ger{\'e}b}, Katinka and {Kilborn}, Virginia and {Wang}, Jing},
        title = "{xGASS: total cold gas scaling relations and molecular-to-atomic gas ratios of galaxies in the local Universe}",
      journal = {\mnras},
         year = 2018,
        month = may,
       volume = {476},
       number = {1},
        pages = {875-895},
          doi = {10.1093/mnras/sty089},
archivePrefix = {arXiv},
       eprint = {1802.02373},
 primaryClass = {astro-ph.GA},
       adsurl = {https://ui.adsabs.harvard.edu/abs/2018MNRAS.476..875C}
}

@ARTICLE{dave17,
       author = {{Dav{\'e}}, Romeel and {Rafieferantsoa}, Mika H. and {Thompson}, Robert J. and {Hopkins}, Philip F.},
        title = "{MUFASA: Galaxy star formation, gas, and metal properties across cosmic time}",
      journal = {\mnras},
         year = 2017,
        month = may,
       volume = {467},
       number = {1},
        pages = {115-132},
          doi = {10.1093/mnras/stx108},
archivePrefix = {arXiv},
       eprint = {1610.01626},
 primaryClass = {astro-ph.GA},
       adsurl = {https://ui.adsabs.harvard.edu/abs/2017MNRAS.467..115D}
}

@ARTICLE{euclid25,
       author = {{Euclid Collaboration} and {Mellier}, Y. and {Abdurro'uf} and {Acevedo Barroso}, J.~A. and {Ach{\'u}carro}, A. and {Adamek}, J. and {Adam}, R. and {Addison}, G.~E. and {Aghanim}, N. and {Aguena}, M. and {Ajani}, V. and {Akrami}, Y. and {Al-Bahlawan}, A. and {Alavi}, A. and {Albuquerque}, I.~S. and {Alestas}, G. and {Alguero}, G. and {Allaoui}, A. and {Allen}, S.~W. and {Allevato}, V. and {Alonso-Tetilla}, A.~V. and {Altieri}, B. and {Alvarez-Candal}, A. and {Alvi}, S. and {Amara}, A. and {Amendola}, L. and {Amiaux}, J. and {Andika}, I.~T. and {Andreon}, S. and {Andrews}, A. and {Angora}, G. and {Angulo}, R.~E. and {Annibali}, F. and {Anselmi}, A. and {Anselmi}, S. and {Arcari}, S. and {Archidiacono}, M. and {Aric{\`o}}, G. and {Arnaud}, M. and {Arnouts}, S. and {Asgari}, M. and {Asorey}, J. and {Atayde}, L. and {Atek}, H. and {Atrio-Barandela}, F. and {Aubert}, M. and {Aubourg}, E. and {Auphan}, T. and {Auricchio}, N. and {Aussel}, B. and {Aussel}, H. and {Avelino}, P.~P. and {Avgoustidis}, A. and {Avila}, S. and {Awan}, S. and {Azzollini}, R. and {Baccigalupi}, C. and {Bachelet}, E. and {Bacon}, D. and {Baes}, M. and {Bagley}, M.~B. and {Bahr-Kalus}, B. and {Balaguera-Antolinez}, A. and {Balbinot}, E. and {Balcells}, M. and {Baldi}, M. and {Baldry}, I. and {Balestra}, A. and {Ballardini}, M. and {Ballester}, O. and {Balogh}, M. and {Ba{\~n}ados}, E. and {Barbier}, R. and {Bardelli}, S. and {Baron}, M. and {Barreiro}, T. and {Barrena}, R. and {Barriere}, J. -C. and {Barros}, B.~J. and {Barthelemy}, A. and {Bartolo}, N. and {Basset}, A. and {Battaglia}, P. and {Battisti}, A.~J. and {Baugh}, C.~M. and {Baumont}, L. and {Bazzanini}, L. and {Beaulieu}, J. -P. and {Beckmann}, V. and {Belikov}, A.~N. and {Bel}, J. and {Bellagamba}, F. and {Bella}, M. and {Bellini}, E. and {Benabed}, K. and {Bender}, R. and {Benevento}, G. and {Bennett}, C.~L. and {Benson}, K. and {Bergamini}, P. and {Bermejo-Climent}, J.~R. and {Bernardeau}, F. and {Bertacca}, D. and {Berthe}, M. and {Berthier}, J. and {Bethermin}, M. and {Beutler}, F. and {Bevillon}, C. and {Bhargava}, S. and {Bhatawdekar}, R. and {Bianchi}, D. and {Bisigello}, L. and {Biviano}, A. and {Blake}, R.~P. and {Blanchard}, A. and {Blazek}, J. and {Blot}, L. and {Bosco}, A. and {Bodendorf}, C. and {Boenke}, T. and {B{\"o}hringer}, H. and {Boldrini}, P. and {Bolzonella}, M. and {Bonchi}, A. and {Bonici}, M. and {Bonino}, D. and {Bonino}, L. and {Bonvin}, C. and {Bon}, W. and {Booth}, J.~T. and {Borgani}, S. and {Borlaff}, A.~S. and {Borsato}, E. and {Bose}, B. and {Botticella}, M.~T. and {Boucaud}, A. and {Bouche}, F. and {Boucher}, J.~S. and {Boutigny}, D. and {Bouvard}, T. and {Bouwens}, R. and {Bouy}, H. and {Bowler}, R.~A.~A. and {Bozza}, V. and {Bozzo}, E. and {Branchini}, E. and {Brando}, G. and {Brau-Nogue}, S. and {Brekke}, P. and {Bremer}, M.~N. and {Brescia}, M. and {Breton}, M. -A. and {Brinchmann}, J. and {Brinckmann}, T. and {Brockley-Blatt}, C. and {Brodwin}, M. and {Brouard}, L. and {Brown}, M.~L. and {Bruton}, S. and {Bucko}, J. and {Buddelmeijer}, H. and {Buenadicha}, G. and {Buitrago}, F. and {Burger}, P. and {Burigana}, C. and {Busillo}, V. and {Busonero}, D. and {Cabanac}, R. and {Cabayol-Garcia}, L. and {Cagliari}, M.~S. and {Caillat}, A. and {Caillat}, L. and {Calabrese}, M. and {Calabro}, A. and {Calderone}, G. and {Calura}, F. and {Camacho Quevedo}, B. and {Camera}, S. and {Campos}, L. and {Ca{\~n}as-Herrera}, G. and {Candini}, G.~P. and {Cantiello}, M. and {Capobianco}, V. and {Cappellaro}, E. and {Cappelluti}, N. and {Cappi}, A. and {Caputi}, K.~I. and {Cara}, C. and {Carbone}, C. and {Cardone}, V.~F. and {Carella}, E. and {Carlberg}, R.~G. and {Carle}, M. and {Carminati}, L. and {Caro}, F. and {Carrasco}, J.~M. and {Carretero}, J. and {Carrilho}, P. and {Carron Duque}, J. and {Carry}, B.},
        title = "{Euclid: I. Overview of the Euclid mission}",
      journal = {\aap},
         year = 2025,
        month = may,
       volume = {697},
          eid = {A1},
        pages = {A1},
          doi = {10.1051/0004-6361/202450810},
archivePrefix = {arXiv},
       eprint = {2405.13491},
 primaryClass = {astro-ph.CO},
       adsurl = {https://ui.adsabs.harvard.edu/abs/2025A&A...697A...1E}
}

@ARTICLE{jwst06,
       author = {{Gardner}, Jonathan P. and {Mather}, John C. and {Clampin}, Mark and {Doyon}, Rene and {Greenhouse}, Matthew A. and {Hammel}, Heidi B. and {Hutchings}, John B. and {Jakobsen}, Peter and {Lilly}, Simon J. and {Long}, Knox S. and {Lunine}, Jonathan I. and {McCaughrean}, Mark J. and {Mountain}, Matt and {Nella}, John and {Rieke}, George H. and {Rieke}, Marcia J. and {Rix}, Hans-Walter and {Smith}, Eric P. and {Sonneborn}, George and {Stiavelli}, Massimo and {Stockman}, H.~S. and {Windhorst}, Rogier A. and {Wright}, Gillian S.},
        title = "{The James Webb Space Telescope}",
      journal = {\ssr},
         year = 2006,
        month = apr,
       volume = {123},
       number = {4},
        pages = {485-606},
          doi = {10.1007/s11214-006-8315-7},
archivePrefix = {arXiv},
       eprint = {astro-ph/0606175},
 primaryClass = {astro-ph},
       adsurl = {https://ui.adsabs.harvard.edu/abs/2006SSRv..123..485G}
}

@ARTICLE{ska09,
       author = {{Dewdney}, P.~E. and {Hall}, P.~J. and {Schilizzi}, R.~T. and {Lazio}, T.~J.~L.~W.},
        title = "{The Square Kilometre Array}",
      journal = {IEEE Proceedings},
         year = 2009,
        month = aug,
       volume = {97},
       number = {8},
        pages = {1482-1496},
          doi = {10.1109/JPROC.2009.2021005},
       adsurl = {https://ui.adsabs.harvard.edu/abs/2009IEEEP..97.1482D}
}

@ARTICLE{mcalpine16,
       author = {{McAlpine}, S. and {Helly}, J.~C. and {Schaller}, M. and {Trayford}, J.~W. and {Qu}, Y. and {Furlong}, M. and {Bower}, R.~G. and {Crain}, R.~A. and {Schaye}, J. and {Theuns}, T. and {Dalla Vecchia}, C. and {Frenk}, C.~S. and {McCarthy}, I.~G. and {Jenkins}, A. and {Rosas-Guevara}, Y. and {White}, S.~D.~M. and {Baes}, M. and {Camps}, P. and {Lemson}, G.},
        title = "{The EAGLE simulations of galaxy formation: Public release of halo and galaxy catalogues}",
      journal = {Astronomy and Computing},
         year = 2016,
        month = apr,
       volume = {15},
        pages = {72-89},
          doi = {10.1016/j.ascom.2016.02.004},
archivePrefix = {arXiv},
       eprint = {1510.01320},
 primaryClass = {astro-ph.GA},
       adsurl = {https://ui.adsabs.harvard.edu/abs/2016A&C....15...72M}
}

@ARTICLE{crain15,
       author = {{Crain}, Robert A. and {Schaye}, Joop and {Bower}, Richard G. and {Furlong}, Michelle and {Schaller}, Matthieu and {Theuns}, Tom and {Dalla Vecchia}, Claudio and {Frenk}, Carlos S. and {McCarthy}, Ian G. and {Helly}, John C. and {Jenkins}, Adrian and {Rosas-Guevara}, Yetli M. and {White}, Simon D.~M. and {Trayford}, James W.},
        title = "{The EAGLE simulations of galaxy formation: calibration of subgrid physics and model variations}",
      journal = {\mnras},
         year = 2015,
        month = jun,
       volume = {450},
       number = {2},
        pages = {1937-1961},
          doi = {10.1093/mnras/stv725},
archivePrefix = {arXiv},
       eprint = {1501.01311},
 primaryClass = {astro-ph.GA},
       adsurl = {https://ui.adsabs.harvard.edu/abs/2015MNRAS.450.1937C}
}

@ARTICLE{planck14,
       author = {{Planck Collaboration} and {Ade}, P.~A.~R. and {Aghanim}, N. and {Alves}, M.~I.~R. and {Armitage-Caplan}, C. and {Arnaud}, M. and {Ashdown}, M. and {Atrio-Barandela}, F. and {Aumont}, J. and {Aussel}, H. and {Baccigalupi}, C. and {Banday}, A.~J. and {Barreiro}, R.~B. and {Barrena}, R. and {Bartelmann}, M. and {Bartlett}, J.~G. and {Bartolo}, N. and {Basak}, S. and {Battaner}, E. and {Battye}, R. and {Benabed}, K. and {Beno{\^\i}t}, A. and {Benoit-L{\'e}vy}, A. and {Bernard}, J. -P. and {Bersanelli}, M. and {Bertincourt}, B. and {Bethermin}, M. and {Bielewicz}, P. and {Bikmaev}, I. and {Blanchard}, A. and {Bobin}, J. and {Bock}, J.~J. and {B{\"o}hringer}, H. and {Bonaldi}, A. and {Bonavera}, L. and {Bond}, J.~R. and {Borrill}, J. and {Bouchet}, F.~R. and {Boulanger}, F. and {Bourdin}, H. and {Bowyer}, J.~W. and {Bridges}, M. and {Brown}, M.~L. and {Bucher}, M. and {Burenin}, R. and {Burigana}, C. and {Butler}, R.~C. and {Calabrese}, E. and {Cappellini}, B. and {Cardoso}, J. -F. and {Carr}, R. and {Carvalho}, P. and {Casale}, M. and {Castex}, G. and {Catalano}, A. and {Challinor}, A. and {Chamballu}, A. and {Chary}, R. -R. and {Chen}, X. and {Chiang}, H.~C. and {Chiang}, L. -Y. and {Chon}, G. and {Christensen}, P.~R. and {Churazov}, E. and {Church}, S. and {Clemens}, M. and {Clements}, D.~L. and {Colombi}, S. and {Colombo}, L.~P.~L. and {Combet}, C. and {Comis}, B. and {Couchot}, F. and {Coulais}, A. and {Crill}, B.~P. and {Cruz}, M. and {Curto}, A. and {Cuttaia}, F. and {Da Silva}, A. and {Dahle}, H. and {Danese}, L. and {Davies}, R.~D. and {Davis}, R.~J. and {de Bernardis}, P. and {de Rosa}, A. and {de Zotti}, G. and {D{\'e}chelette}, T. and {Delabrouille}, J. and {Delouis}, J. -M. and {D{\'e}mocl{\`e}s}, J. and {D{\'e}sert}, F. -X. and {Dick}, J. and {Dickinson}, C. and {Diego}, J.~M. and {Dolag}, K. and {Dole}, H. and {Donzelli}, S. and {Dor{\'e}}, O. and {Douspis}, M. and {Ducout}, A. and {Dunkley}, J. and {Dupac}, X. and {Efstathiou}, G. and {Elsner}, F. and {En{\ss}lin}, T.~A. and {Eriksen}, H.~K. and {Fabre}, O. and {Falgarone}, E. and {Falvella}, M.~C. and {Fantaye}, Y. and {Fergusson}, J. and {Filliard}, C. and {Finelli}, F. and {Flores-Cacho}, I. and {Foley}, S. and {Forni}, O. and {Fosalba}, P. and {Frailis}, M. and {Fraisse}, A.~A. and {Franceschi}, E. and {Freschi}, M. and {Fromenteau}, S. and {Frommert}, M. and {Gaier}, T.~C. and {Galeotta}, S. and {Gallegos}, J. and {Galli}, S. and {Gandolfo}, B. and {Ganga}, K. and {Gauthier}, C. and {G{\'e}nova-Santos}, R.~T. and {Ghosh}, T. and {Giard}, M. and {Giardino}, G. and {Gilfanov}, M. and {Girard}, D. and {Giraud-H{\'e}raud}, Y. and {Gjerl{\o}w}, E. and {Gonz{\'a}lez-Nuevo}, J. and {G{\'o}rski}, K.~M. and {Gratton}, S. and {Gregorio}, A. and {Gruppuso}, A. and {Gudmundsson}, J.~E. and {Haissinski}, J. and {Hamann}, J. and {Hansen}, F.~K. and {Hansen}, M. and {Hanson}, D. and {Harrison}, D.~L. and {Heavens}, A. and {Helou}, G. and {Hempel}, A. and {Henrot-Versill{\'e}}, S. and {Hern{\'a}ndez-Monteagudo}, C. and {Herranz}, D. and {Hildebrandt}, S.~R. and {Hivon}, E. and {Ho}, S. and {Hobson}, M. and {Holmes}, W.~A. and {Hornstrup}, A. and {Hou}, Z. and {Hovest}, W. and {Huey}, G. and {Huffenberger}, K.~M. and {Hurier}, G. and {Ili{\'c}}, S. and {Jaffe}, A.~H. and {Jaffe}, T.~R. and {Jasche}, J. and {Jewell}, J. and {Jones}, W.~C. and {Juvela}, M. and {Kalberla}, P. and {Kangaslahti}, P. and {Keih{\"a}nen}, E. and {Kerp}, J. and {Keskitalo}, R. and {Khamitov}, I. and {Kiiveri}, K. and {Kim}, J. and {Kisner}, T.~S. and {Kneissl}, R. and {Knoche}, J. and {Knox}, L. and {Kunz}, M. and {Kurki-Suonio}, H. and {Lacasa}, F. and {Lagache}, G. and {L{\"a}hteenm{\"a}ki}, A. and {Lamarre}, J. -M. and {Langer}, M. and {Lasenby}, A. and {Lattanzi}, M. and {Laureijs}, R.~J. and {Lavabre}, A. and {Lawrence}, C.~R. and {Le Jeune}, M. and {Leach}, S. and {Leahy}, J.~P.},
        title = "{Planck 2013 results. I. Overview of products and scientific results}",
      journal = {\aap},
         year = 2014,
        month = nov,
       volume = {571},
          eid = {A1},
        pages = {A1},
          doi = {10.1051/0004-6361/201321529},
archivePrefix = {arXiv},
       eprint = {1303.5062},
 primaryClass = {astro-ph.CO},
       adsurl = {https://ui.adsabs.harvard.edu/abs/2014A&A...571A...1P}
}

@ARTICLE{trayford15,
       author = {{Trayford}, James W. and {Theuns}, Tom and {Bower}, Richard G. and {Schaye}, Joop and {Furlong}, Michelle and {Schaller}, Matthieu and {Frenk}, Carlos S. and {Crain}, Robert A. and {Dalla Vecchia}, Claudio and {McCarthy}, Ian G.},
        title = "{Colours and luminosities of z = 0.1 galaxies in the EAGLE simulation}",
      journal = {\mnras},
         year = 2015,
        month = sep,
       volume = {452},
       number = {3},
        pages = {2879-2896},
          doi = {10.1093/mnras/stv1461},
archivePrefix = {arXiv},
       eprint = {1504.04374},
 primaryClass = {astro-ph.GA},
       adsurl = {https://ui.adsabs.harvard.edu/abs/2015MNRAS.452.2879T}
}

@ARTICLE{doi10,
       author = {{Doi}, Mamoru and {Tanaka}, Masayuki and {Fukugita}, Masataka and {Gunn}, James E. and {Yasuda}, Naoki and {Ivezi{\'c}}, {\v{Z}}eljko and {Brinkmann}, Jon and {de Haars}, Ernst and {Kleinman}, S.~J. and {Krzesinski}, Jurek and {French Leger}, R.},
        title = "{Photometric Response Functions of the Sloan Digital Sky Survey Imager}",
      journal = {\aj},
         year = 2010,
        month = apr,
       volume = {139},
       number = {4},
        pages = {1628-1648},
          doi = {10.1088/0004-6256/139/4/1628},
archivePrefix = {arXiv},
       eprint = {1002.3701},
 primaryClass = {astro-ph.IM},
       adsurl = {https://ui.adsabs.harvard.edu/abs/2010AJ....139.1628D}
}

@ARTICLE{casertano85,
       author = {{Casertano}, S. and {Hut}, P.},
        title = "{Core radius and density measurements in N-body experiments Connections with theoretical and observational definitions}",
      journal = {\apj},
         year = 1985,
        month = nov,
       volume = {298},
        pages = {80-94},
          doi = {10.1086/163589},
       adsurl = {https://ui.adsabs.harvard.edu/abs/1985ApJ...298...80C}
}

@ARTICLE{baldwin81,
       author = {{Baldwin}, J.~A. and {Phillips}, M.~M. and {Terlevich}, R.},
        title = "{Classification parameters for the emission-line spectra of extragalactic objects.}",
      journal = {\pasp},
         year = 1981,
        month = feb,
       volume = {93},
        pages = {5-19},
          doi = {10.1086/130766},
       adsurl = {https://ui.adsabs.harvard.edu/abs/1981PASP...93....5B}
}

@ARTICLE{mcalpine20,
       author = {{McAlpine}, Stuart and {Harrison}, Chris M. and {Rosario}, David J. and {Alexander}, David M. and {Ellison}, Sara L. and {Johansson}, Peter H. and {Patton}, David R.},
        title = "{Galaxy mergers in EAGLE do not induce a significant amount of black hole growth yet do increase the rate of luminous AGN}",
      journal = {\mnras},
         year = 2020,
        month = jun,
       volume = {494},
       number = {4},
        pages = {5713-5733},
          doi = {10.1093/mnras/staa1123},
archivePrefix = {arXiv},
       eprint = {2002.00959},
 primaryClass = {astro-ph.GA},
       adsurl = {https://ui.adsabs.harvard.edu/abs/2020MNRAS.494.5713M}
}

@ARTICLE{kauffmann03,
       author = {{Kauffmann}, Guinevere and {Heckman}, Timothy M. and {White}, Simon D.~M. and {Charlot}, St{\'e}phane and {Tremonti}, Christy and {Peng}, Eric W. and {Seibert}, Mark and {Brinkmann}, Jon and {Nichol}, Robert C. and {SubbaRao}, Mark and {York}, Don},
        title = "{The dependence of star formation history and internal structure on stellar mass for {}10$^{5}$ low-redshift galaxies}",
      journal = {\mnras},
         year = 2003,
        month = may,
       volume = {341},
       number = {1},
        pages = {54-69},
          doi = {10.1046/j.1365-8711.2003.06292.x},
archivePrefix = {arXiv},
       eprint = {astro-ph/0205070},
 primaryClass = {astro-ph},
       adsurl = {https://ui.adsabs.harvard.edu/abs/2003MNRAS.341...54K}
}

@ARTICLE{binney04,
       author = {{Binney}, James},
        title = "{On the origin of the galaxy luminosity function}",
      journal = {\mnras},
         year = 2004,
        month = feb,
       volume = {347},
       number = {4},
        pages = {1093-1096},
          doi = {10.1111/j.1365-2966.2004.07277.x},
archivePrefix = {arXiv},
       eprint = {astro-ph/0308172},
 primaryClass = {astro-ph},
       adsurl = {https://ui.adsabs.harvard.edu/abs/2004MNRAS.347.1093B}
}

@ARTICLE{birnboim03,
       author = {{Birnboim}, Yuval and {Dekel}, Avishai},
        title = "{Virial shocks in galactic haloes?}",
      journal = {\mnras},
         year = 2003,
        month = oct,
       volume = {345},
       number = {1},
        pages = {349-364},
          doi = {10.1046/j.1365-8711.2003.06955.x},
archivePrefix = {arXiv},
       eprint = {astro-ph/0302161},
 primaryClass = {astro-ph},
       adsurl = {https://ui.adsabs.harvard.edu/abs/2003MNRAS.345..349B}
}

@ARTICLE{dekel06,
       author = {{Dekel}, Avishai and {Birnboim}, Yuval},
        title = "{Galaxy bimodality due to cold flows and shock heating}",
      journal = {\mnras},
         year = 2006,
        month = may,
       volume = {368},
       number = {1},
        pages = {2-20},
          doi = {10.1111/j.1365-2966.2006.10145.x},
archivePrefix = {arXiv},
       eprint = {astro-ph/0412300},
 primaryClass = {astro-ph},
       adsurl = {https://ui.adsabs.harvard.edu/abs/2006MNRAS.368....2D}
}

@ARTICLE{keres05,
       author = {{Kere{\v{s}}}, Du{\v{s}}an and {Katz}, Neal and {Weinberg}, David H. and {Dav{\'e}}, Romeel},
        title = "{How do galaxies get their gas?}",
      journal = {\mnras},
         year = 2005,
        month = oct,
       volume = {363},
       number = {1},
        pages = {2-28},
          doi = {10.1111/j.1365-2966.2005.09451.x},
archivePrefix = {arXiv},
       eprint = {astro-ph/0407095},
 primaryClass = {astro-ph},
       adsurl = {https://ui.adsabs.harvard.edu/abs/2005MNRAS.363....2K}
}

@ARTICLE{gabor10,
       author = {{Gabor}, J.~M. and {Dav{\'e}}, R. and {Finlator}, K. and {Oppenheimer}, B.~D.},
        title = "{How is star formation quenched in massive galaxies?}",
      journal = {\mnras},
         year = 2010,
        month = sep,
       volume = {407},
       number = {2},
        pages = {749-771},
          doi = {10.1111/j.1365-2966.2010.16961.x},
archivePrefix = {arXiv},
       eprint = {1001.1734},
 primaryClass = {astro-ph.CO},
       adsurl = {https://ui.adsabs.harvard.edu/abs/2010MNRAS.407..749G}
}

@ARTICLE{gabor15,
       author = {{Gabor}, J.~M. and {Dav{\'e}}, R.},
        title = "{Hot gas in massive haloes drives both mass quenching and environment quenching}",
      journal = {\mnras},
         year = 2015,
        month = feb,
       volume = {447},
       number = {1},
        pages = {374-391},
          doi = {10.1093/mnras/stu2399},
archivePrefix = {arXiv},
       eprint = {1405.1043},
 primaryClass = {astro-ph.GA},
       adsurl = {https://ui.adsabs.harvard.edu/abs/2015MNRAS.447..374G}
}

@ARTICLE{patton11,
       author = {{Patton}, David R. and {Ellison}, Sara L. and {Simard}, Luc and {McConnachie}, Alan W. and {Mendel}, J. Trevor},
        title = "{Galaxy pairs in the Sloan Digital Sky Survey - III. Evidence of induced star formation from optical colours}",
      journal = {\mnras},
         year = 2011,
        month = mar,
       volume = {412},
       number = {1},
        pages = {591-606},
          doi = {10.1111/j.1365-2966.2010.17932.x},
archivePrefix = {arXiv},
       eprint = {1010.5778},
 primaryClass = {astro-ph.CO},
       adsurl = {https://ui.adsabs.harvard.edu/abs/2011MNRAS.412..591P}
}

@ARTICLE{shakura73,
       author = {{Shakura}, N.~I. and {Sunyaev}, R.~A.},
        title = "{Black holes in binary systems. Observational appearance.}",
      journal = {\aap},
         year = 1973,
        month = jan,
       volume = {24},
        pages = {337-355},
       adsurl = {https://ui.adsabs.harvard.edu/abs/1973A&A....24..337S}
}

@ARTICLE{furlong15,
       author = {{Furlong}, M. and {Bower}, R.~G. and {Theuns}, T. and {Schaye}, J. and {Crain}, R.~A. and {Schaller}, M. and {Dalla Vecchia}, C. and {Frenk}, C.~S. and {McCarthy}, I.~G. and {Helly}, J. and {Jenkins}, A. and {Rosas-Guevara}, Y.~M.},
        title = "{Evolution of galaxy stellar masses and star formation rates in the EAGLE simulations}",
      journal = {\mnras},
         year = 2015,
        month = jul,
       volume = {450},
       number = {4},
        pages = {4486-4504},
          doi = {10.1093/mnras/stv852},
archivePrefix = {arXiv},
       eprint = {1410.3485},
 primaryClass = {astro-ph.GA},
       adsurl = {https://ui.adsabs.harvard.edu/abs/2015MNRAS.450.4486F}
}

@ARTICLE{baes20,
       author = {{Baes}, Maarten and {Tr{\v{c}}ka}, Ana and {Camps}, Peter and {Trayford}, James and {Katsianis}, Antonios and {Marchetti}, Lucia and {Theuns}, Tom and {Vaccari}, Mattia and {Vandenbroucke}, Bert},
        title = "{Infrared luminosity functions and dust mass functions in the EAGLE simulation}",
      journal = {\mnras},
         year = 2020,
        month = may,
       volume = {494},
       number = {2},
        pages = {2912-2924},
          doi = {10.1093/mnras/staa990},
archivePrefix = {arXiv},
       eprint = {2004.02713},
 primaryClass = {astro-ph.GA},
       adsurl = {https://ui.adsabs.harvard.edu/abs/2020MNRAS.494.2912B}
}

@ARTICLE{katsianis17,
       author = {{Katsianis}, A. and {Blanc}, G. and {Lagos}, C.~P. and {Tejos}, N. and {Bower}, R.~G. and {Alavi}, A. and {Gonzalez}, V. and {Theuns}, T. and {Schaller}, M. and {Lopez}, S.},
        title = "{The evolution of the star formation rate function in the EAGLE simulations: a comparison with UV, IR and H{\ensuremath{\alpha}} observations from z {\ensuremath{\sim}} 8 to z {\ensuremath{\sim}} 0}",
      journal = {\mnras},
         year = 2017,
        month = nov,
       volume = {472},
       number = {1},
        pages = {919-939},
          doi = {10.1093/mnras/stx2020},
archivePrefix = {arXiv},
       eprint = {1708.01913},
 primaryClass = {astro-ph.GA},
       adsurl = {https://ui.adsabs.harvard.edu/abs/2017MNRAS.472..919K}
}

@ARTICLE{vellis17,
       author = {{Velliscig}, Marco and {Cacciato}, Marcello and {Hoekstra}, Henk and {Schaye}, Joop and {Heymans}, Catherine and {Hildebrandt}, Hendrik and {Loveday}, Jon and {Norberg}, Peder and {Sif{\'o}n}, Crist{\'o}bal and {Schneider}, Peter and {van Uitert}, Edo and {Viola}, Massimo and {Brough}, Sarah and {Erben}, Thomas and {Holwerda}, Benne W. and {Hopkins}, Andrew M. and {Kuijken}, Konrad},
        title = "{Galaxy-galaxy lensing in EAGLE: comparison with data from 180 deg$^{2}$ of the KiDS and GAMA surveys}",
      journal = {\mnras},
         year = 2017,
        month = nov,
       volume = {471},
       number = {3},
        pages = {2856-2870},
          doi = {10.1093/mnras/stx1789},
archivePrefix = {arXiv},
       eprint = {1612.04825},
 primaryClass = {astro-ph.GA},
       adsurl = {https://ui.adsabs.harvard.edu/abs/2017MNRAS.471.2856V}
}

@ARTICLE{guevara16,
       author = {{Rosas-Guevara}, Yetli and {Bower}, Richard G. and {Schaye}, Joop and {McAlpine}, Stuart and {Dalla Vecchia}, Claudio and {Frenk}, Carlos S. and {Schaller}, Matthieu and {Theuns}, Tom},
        title = "{Supermassive black holes in the EAGLE Universe. Revealing the observables of their growth}",
      journal = {\mnras},
         year = 2016,
        month = oct,
       volume = {462},
       number = {1},
        pages = {190-205},
          doi = {10.1093/mnras/stw1679},
archivePrefix = {arXiv},
       eprint = {1604.00020},
 primaryClass = {astro-ph.GA},
       adsurl = {https://ui.adsabs.harvard.edu/abs/2016MNRAS.462..190R}
}

@ARTICLE{gunn72,
       author = {{Gunn}, James E. and {Gott}, J. Richard, III},
        title = "{On the Infall of Matter Into Clusters of Galaxies and Some Effects on Their Evolution}",
      journal = {\apj},
         year = 1972,
        month = aug,
       volume = {176},
        pages = {1},
          doi = {10.1086/151605},
       adsurl = {https://ui.adsabs.harvard.edu/abs/1972ApJ...176....1G}
}

@ARTICLE{moore96,
       author = {{Moore}, Ben and {Katz}, Neal and {Lake}, George and {Dressler}, Alan and {Oemler}, Augustus},
        title = "{Galaxy harassment and the evolution of clusters of galaxies}",
      journal = {\nat},
         year = 1996,
        month = feb,
       volume = {379},
       number = {6566},
        pages = {613-616},
          doi = {10.1038/379613a0},
archivePrefix = {arXiv},
       eprint = {astro-ph/9510034},
 primaryClass = {astro-ph},
       adsurl = {https://ui.adsabs.harvard.edu/abs/1996Natur.379..613M}
}

@ARTICLE{balogh00,
       author = {{Balogh}, Michael L. and {Navarro}, Julio F. and {Morris}, Simon L.},
        title = "{The Origin of Star Formation Gradients in Rich Galaxy Clusters}",
      journal = {\apj},
         year = 2000,
        month = sep,
       volume = {540},
       number = {1},
        pages = {113-121},
          doi = {10.1086/309323},
archivePrefix = {arXiv},
       eprint = {astro-ph/0004078},
 primaryClass = {astro-ph},
       adsurl = {https://ui.adsabs.harvard.edu/abs/2000ApJ...540..113B}
}

@ARTICLE{larson80,
       author = {{Larson}, R.~B. and {Tinsley}, B.~M. and {Caldwell}, C.~N.},
        title = "{The evolution of disk galaxies and the origin of S0 galaxies}",
      journal = {\apj},
         year = 1980,
        month = may,
       volume = {237},
        pages = {692-707},
          doi = {10.1086/157917},
       adsurl = {https://ui.adsabs.harvard.edu/abs/1980ApJ...237..692L}
}

\end{document}